\documentclass[12pt]{article}
\usepackage{amscd}
\usepackage{verbatim}
\usepackage{amssymb}
\usepackage{amsmath}
\usepackage{tabularx}
\usepackage{threeparttable}
\newcommand{\be}{\begin{equation}}
\newcommand{\ee}{\end{equation}}
\newcommand{\ben}{\begin{eqnarray}}
\newcommand{\een}{\end{eqnarray}}
\newcommand{\ba}{\begin{eqnarray}}
\newcommand{\ea}{\end{eqnarray}}

\newcommand{\bi}{\begin{itemize}}
\newcommand{\ei}{\end{itemize}}

\begin{document}
\begin{center}
\vspace{24pt} { \large \bf Gravitational Perturbations in a cavity: Nonlinearities} \\
\vspace{30pt}
\vspace{30pt}
\vspace{30pt}
{\bf Dhanya S.Menon\footnote{dhanya.menon@students.iiserpune.ac.in}}, {\bf Vardarajan
Suneeta\footnote{suneeta@iiserpune.ac.in}}\\
\vspace{24pt} %%
{\em  The Indian Institute of Science Education and Research (IISER),\\
Pune, India - 411008.}
\end{center}
\date{\today}
\bigskip
\begin{center}
{\bf Abstract}
\end{center}
Motivated by recent studies of nonlinear perturbations of asymptotically anti-de Sitter spacetimes, we study gravitational perturbations of $(n+2)$ dimensional Minkowski spacetime
with a spherical Dirichlet wall. By considering the tensor, vector and scalar perturbations on the $n$ sphere, we present simplified nonhomogeneous equations at arbitrary order in weakly nonlinear
perturbation theory for each sector. A suitable choice of perturbative variables is required at higher orders to simplify the expression for the boundary conditions and to expand the variables in terms of linear order eigenfunctions.
Finally we comment on the nonlinear stability of the system. Some of the tools used can easily be generalized to study nonlinear
perturbations of anti-de Sitter spacetime.

\newpage
\section{Introduction}
The seminal work of Bizon and Rostworowski
\cite{bizon} demonstrated that four dimensional anti-de Sitter spacetime ($AdS$) was
nonlinearly unstable to spherically
symmetric massless scalar field perturbations. The end-point of the
instability
was a Schwarzschild-$AdS$ black hole.\footnote {Further, it was shown that the Schwarzchild-$AdS$ black hole was
stable for the spherically symmetric Einstein-Klein-Gordon system \cite{holzegel1}.} It was thus concluded that $AdS$ was unstable
to black hole formation for a large class of arbitrarily small
perturbations and that reflecting boundary conditions played a key role in
causing instability \cite{bizon1}.
Later, it was demonstrated in
\cite{jalmuzna} that the instability was seen in all dimensions.
 The instability was also present for
complex scalar fields \cite{liebling1} in $AdS$
spacetime. The necessary conditions for an $AdS$-like instability were analyzed in
\cite{ads}. In \cite{oliveira}, turbulent behaviour characterized by a Kolmogorov-Zakharov power
spectrum was uncovered for the Klein-Gordon gravity system.
Non-collapsing solutions in asymptotically $AdS$ spacetimes were studied in \cite{maliborski1}, \cite{buchel}, \cite{maliborski5}, \cite{arias}, \cite{kim}, \cite{fodor}, \cite{maliborski3}. Going beyond Einstein gravity, the system of a scalar field and gravity with a Gauss-Bonnet term was analyzed in \cite{kunstatter}, \cite{kunstatter1}. The $AdS$ instability problem
was also studied in \cite{balas}, \cite{green}, \cite{maliborski6}, using the two-time framework (TTF) and a careful analysis using rigorous renormalization group methods \cite{craps1}, \cite{craps2}. Interacting scalar fields in $AdS$ were studied in \cite{basu}.
Of particular interest are numerical studies of non-spherically symmetric collapse in the Einstein
gravity-scalar field system in asymptotically $AdS$ spacetime
\cite{bantilan},
\cite{choptuik}. A massless scalar field in flat space
enclosed in a spherical cavity was studied as a toy model for $AdS$-like
boundary conditions and it was shown
to lead to a nonlinear instability \cite{maliborski}.
The scalar field-gravity system in a cavity in flat space with self-interactions of the field was
studied in
\cite{pani}.
A comprehensive review of work on the instability of
$AdS$, particularly the scalar field-gravity system can be found in \cite{martinonreview}.

Gravitational turbulent instability was first studied in $AdS$ in \cite{horowitz}.
This uncovered geons --- time-periodic,
asymptotically $AdS$ solutions that were stable \cite{horowitz}
see also \cite{horowitz1}, \cite{dias}, \cite{rost}, \cite{rost2}, \cite{diasantos}, \cite{martinon2},
\cite{rost3}, \cite{fodor1}. Purely gravitational perturbations of $AdS$ satisfying the
cohomogeneity-two biaxial Bianchi IX ansatz were studied and black hole
formation was observed in \cite{cohomo}.

In this work, we depart from the system of gravitational perturbations of $AdS$ spacetime. Instead,
we consider gravitational perturbations of Minkowski spacetime with a spherical Dirichlet wall
in $(n+2)$ dimensions with $n \ge 2$. The box-like boundary conditions mimic those in $AdS$ spacetime --- however, we find important differences both from the point of view of stability and in the mathematical analysis of higher order equations.
Linearized perturbations of this system have been studied in
\cite{marolf} and it has been shown to be stable. Further, linear perturbations have an asymptotically resonant spectrum. We extend this
study to the nonlinear regime using weakly nonlinear perturbation theory. Our analysis shows that the system is nonlinearly
stable under arbitrarily small perturbations. An indicator of the magnitude of the perturbation that may be required to trigger instability is the deviation of this asymptotically resonant spectrum from a fully resonant one. This may be seen in numerical studies.

In order to
write down the perturbation equations, we use the gauge invariant formalism
developed by Kodama, Ishibashi and Seto \cite{seto} which we extend to higher order. We present simplified equations for the
tensor, vector and scalar (on the $n$ sphere) perturbations.
In all cases, there are subtleties involved in the imposition of Dirichlet wall boundary conditions at higher orders, and in the
analysis of solutions to the perturbation equations satisfying these boundary conditions. We analyze these solutions at arbitrary order and apply the results in
\cite{ads} to comment on the nonlinear stability of this system. Section II describes the methodology of the perturbation theory analysis at arbitrary order. A brief summary of the Ishibashi-Kodama-Seto formalism is also given. The linearized tensor, vector and scalar perturbations are analyzed in section III. The scalar eigenfunctions at linear order satisfy a modified orthogonality relation owing to the appearance of frequency dependent boundary conditions in the scalar sector. Section IV contains the higher order perturbation equations. By defining shifted perturbation variables (shifted by source terms) at higher order when necessary, we can expand the perturbation variables in terms of the linear order eigenfunctions, with time dependent coefficients
obeying a forced harmonic oscillator equation. Section V is a brief section on how the various source terms are obtained from the solutions to perturbation equations at lower orders. In section VI, we analyze special modes with $l_s = 0,1$ in the case of scalar modes and $l_v = 1$ for vector modes. For these modes, the perturbation equations need appropriate gauge fixing. The modes can be obtained by solving these equations and it can be readily seen that at linear order, these modes reduce to pure gauge as expected.
Section VII contains a summary and discussion of the nonlinear stability of this system, by using the equations obtained in the paper and nonlinear dynamics results in \cite{ads}. Appendices A and B have detailed computations for some of the equations in the rest of the paper. Appendix C contains an explicit evaluation of the second order source terms in the case when we have only tensor perturbations at linear order.
\section{Methodology}
We use perturbation theory to study gravitational perturbations in $(n+2)$ dimensional Minkowski spacetime with a spherical Dirichlet wall. For this we need to find solutions to the Einstein's field equations with $\Lambda=0$.
\begin{align}
G_{\mu\nu}=R_{\mu\nu}-\frac{1}{2}g_{\mu\nu}R=0.
\label{eq1}
\end{align}
We now follow the analysis and partly the notation in Rostworowski \cite{rost}. The `bar' quantities refer to the background  Minkowski metric $ds^2= - dt^2+dr^2+r^2d\Omega^2_{n}$,
where $d\Omega^2_{n}$ is the metric for the $n-$sphere, $n \ge 2$.
Since we are dealing with weakly nonlinear perturbations, we let $g_{\mu\nu}=\bar{g}_{\mu\nu}+\delta g_{\mu\nu}$ where
\begin{align}
\delta g_{\mu\nu}=\sum_{1\leq i}{^{(i)}} h_{\mu\nu}\epsilon^i.
\label{eq3}
\end{align}
Then the inverse metric is given by
\begin{align}
g^{\alpha\beta}&=(\bar{g}^{-1}-\bar{g}^{-1}\delta g \bar{g}^{-1}+\bar{g}^{-1}\delta g \bar{g}^{-1}\delta g \bar{g}^{-1}-...)^{\alpha\beta}\nonumber\\&=\bar{g}^{\alpha\beta}+\delta g^{\alpha\beta}.
\label{eq4}
\end{align}
 The Christoffel symbol is decomposed as
\begin{align}
\Gamma^{\alpha}_{\mu\nu}&=\bar{\Gamma}^{\alpha}_{\mu\nu}+\frac{1}{2}(\bar{g}^{-1}-\bar{g}^{-1}\delta g \bar{g}^{-1}+...)^{\alpha\beta}( \bar{\nabla}_{\mu}\delta g_{\beta\nu}+\bar{\nabla}_{\nu}\delta g_{\beta\mu}-\bar{\nabla}_{\beta}\delta g_{\mu\nu} )\nonumber\\&=\bar{\Gamma}^{\alpha}_{\mu\nu}+\delta \Gamma^{\alpha}_{\mu\nu}.
\label{eq5}
\end{align}
Similarly the Ricci tensor is decomposed as
\begin{align}
R_{\mu\nu}&=\bar{R}_{\mu\nu}+\bar{\nabla}_{\alpha}\delta \Gamma^{\alpha}_{\mu\nu}-\bar{\nabla}_{\nu}\delta \Gamma_{\alpha\mu}^{\alpha}+\delta\Gamma^{\alpha}_{\alpha\lambda}\delta \Gamma^{\lambda}_{\mu\nu}-\delta \Gamma^{\lambda}_{\mu\alpha}\delta \Gamma^{\alpha}_{\lambda\nu}\nonumber\\&=\bar{R}_{\mu\nu}+\delta R_{\mu\nu}.
\label{eq6}
\end{align}
  The Lorentzian Lichnerowicz operator $\triangle_L$ is defined as
\begin{align}
2\triangle_L {^{(i)}}h_{\mu\nu}=&-\bar{\nabla}^{\alpha}\bar{\nabla}_{\alpha}{^{(i)}}h_{\mu\nu}-\bar{\nabla}_{\mu}\bar{\nabla}_{\nu}{^{(i)}}h+\bar{\nabla}_{\mu}\bar{\nabla}_{\alpha}{^{(i)}}h^{\alpha}_{\nu}+\bar{\nabla}_{\nu}\bar{\nabla}_{\alpha}{^{(i)}}h^{\alpha}_{\mu}\nonumber\\&+\bar{R}_{\mu\alpha}{^{(i)}}\!h^{\alpha}_{\nu}+\bar{R}_{\nu\alpha}{^{(i)}}\!h^{\alpha}_{\mu}-2\bar{R}_{\mu\alpha\nu\lambda}{^{(i)}}h^{\alpha\lambda}.\label{eq6.3}
\end{align}
We also define a quantity ${^{(i)}}A_{\mu\nu}$ as
\begin{align}
{^{(i)}}\!A_{\mu\nu}=&[\epsilon^i]\Big\{-\bar{\nabla}_{\alpha}\big[ (-\bar{g}^{-1}\delta g\bar{g}^{-1}+...)^{\alpha\lambda}(\bar{\nabla}_{\mu}\delta g_{\lambda\nu}+\bar{\nabla}_{\nu}\delta g_{\lambda\nu}-\bar{\nabla}_{\lambda}\delta g_{\mu\nu})\big] \nonumber\\&+\bar{\nabla}_{\nu}\big[ (-\bar{g}^{-1}\delta g\bar{g}^{-1}+...)^{\alpha\lambda}(\bar{\nabla}_{\mu}\delta g_{\lambda\alpha}+\bar{\nabla}_{\alpha}\delta g_{\lambda\mu}-\bar{\nabla}_{\lambda}\delta g_{\mu\alpha})\big]\nonumber\\&-
2\delta\Gamma^{\alpha}_{\alpha\lambda}\delta\Gamma^{\lambda}_{\mu\nu}+2\delta\Gamma^{\lambda}_{\mu\alpha}\delta\Gamma^{\alpha}_{\lambda\nu}  \Big\},
\label{eq6.5}
\end{align}
where $[\epsilon^i] \; f $ denotes the coefficient of $\epsilon^i$ in the expansion of power series $\sum_{i}\epsilon^i f_i$. Since the background metric is Minkowski, the Ricci and the Riemann tensor, $\bar{R}_{\mu\nu}$ and $\bar{R}^{\mu}_{\:\nu\alpha\beta}$ vanish. Moreover, the total metric  $g_{\mu\nu}$  is the solution of the vacuum Einstein's equation (\ref{eq1}). Therefore, by plugging the expressions (\ref{eq4}), (\ref{eq5}) and (\ref{eq6}) in
(\ref{eq1}) and collecting terms in like powers of $\epsilon$, we
obtain
\begin{align}
2\tilde{\triangle}_L{^{(i)}}\!h_{\mu\nu}={^{(i)}}\!S_{\mu\nu},
\label{eq6.1}
\end{align}
where $\tilde{\triangle}_L{^{(i)}}\!h_{\mu\nu}$ is defined as
\begin{align}
2\tilde{\triangle}_L{^{(i)}}\!h_{\mu\nu}=2\triangle_L{^{(i)}}\!h_{\mu\nu}-\bar{g}_{\mu\nu}\bar{g}^{\alpha\beta}\triangle_L{^{(i)}}\!h_{\alpha\beta}
\label{eq6.2}
\end{align}
and the source ${^{(i)}}\!S_{\mu\nu}$ is given in terms of ${^{(i)}}\!A_{\mu\nu}$ as
\begin{align}
{^{(i)}}\!S_{\mu\nu}={^{(i)}}\!A_{\mu\nu}-\frac{1}{2}\bar{g}_{\mu\nu}\bar{g}^{\alpha\beta}{^{(i)}}\!A_{\alpha\beta}.
\label{eq6.4}
\end{align}
 The background Minkowski  metric $\bar{g}_{\mu\nu}$ is of the following form
\begin{align}
ds^2=\bar{g}_{\mu\nu}dz^{\mu}dz^{\nu}=g_{ab}(y)dy^ady^b+r^2(y)d\Omega_n^2,
\label{eq7}
\end{align}
where the metric $d\Omega^2_n$
\begin{align}
d\Omega_n^2=\gamma_{ij}(w)dw^idw^j,
\label{eq8}
\end{align}
is the  metric for the $n-$sphere \& has a constant sectional curvature $K=1$.

One can use the gauge invariant formalism of Ishibashi, Kodama and Seto \cite{seto} to study the perturbations, the difference being that we extend it to higher orders as well.
We associate a covariant derivative each with $ds^2$, $g_{ab}dy^ady^b$ and $d\Omega_n^2$ which are $\bar{\nabla}_M$, $\bar{D}_a$ and  $\bar{D}_i$ respectively. We will also decompose the metric perturbations ${^{(i)}}h_{\mu\nu}$ according to their behaviour on the $n-$sphere i.e. into the  scalar type, $\mathbb{S}$, the vector type, $\mathbb{V}_i$  and the tensor type, $\mathbb{T}_{ij}$.
Note that in the following sections, $\hat{\vartriangle}=\hat{D}^i\hat{D}_i$ where raising (and lowering) of $\hat{D}_i$ (and $\hat{D}^i$) is done with $\gamma_{ij}$.
The scalar harmonics $\mathbb{S}$ satisfy
\begin{align}
(\hat{\vartriangle}+k^2_s)\mathbb{S}=0,
\label{eq9}
\end{align}
with $k_s^2=l_s(l_s+n-1)$ where $l_s=0,1,...$ from where one can construct scalar type vector harmonics $\mathbb{S}_i$
\begin{align}
\mathbb{S}_i=-\frac{1}{k_s}\bar{D}_i\mathbb{S}
\label{eq10}
\end{align}
 which satisfy
\begin{align}
\bar{D}^i\mathbb{S}_i=k_s\mathbb{S}
\label{eq11}
\end{align}
as well as scalar type tensor harmonics
\begin{align}
\mathbb{S}_{ij}=\frac{1}{k^2_s}\bar{D}_i\bar{D}_j\mathbb{S}+\frac{1}{n}\gamma_{ij}\mathbb{S}
\label{eq12}
\end{align}
which satisfy
\begin{align}
{\mathbb{S}_i^i=0;~~\bar{D}_j\mathbb{S}^j_i=\frac{(n-1)(k^2_s-nK)}{nk_s}\mathbb{S}_i}.
\label{eq13}
\end{align}
Vector harmonics $\mathbb{V}_i$ are defined by
\begin{align}
(\hat{\vartriangle}+k^2_v)\mathbb{V}_i=0
\label{eq14}
\end{align}
with $k^2_v=l_v(l_v+n-1)-1$ where $l_v=1,2,...$ such that
\begin{align}
{\bar{D}_i\mathbb{V}^i=0}.
\label{eq15}
\end{align}
One can construct the following tensor from the vector harmonics:
\begin{align}
\mathbb{V}_{ij}=-\frac{1}{2k_v}(\bar{D}_i\mathbb{V}_j+\bar{D}_j\mathbb{V}_i),
\label{eq16}
\end{align}
 which satisfy
\begin{align}
\mathbb{V}_i^i=0;~~ \bar{D}_j\mathbb{V}_i^j=\frac{(k^2_v-(n-1)K)}{2k_v}\mathbb{V}_i.
\label{eq17}
\end{align}
Tensor harmonics, $\mathbb{T}_{ij}$ are defined for $n>2$ by
\begin{align}
(\hat{\vartriangle}+k^2)\mathbb{T}_{ij}=0
\label{eq18}
\end{align}
with $k^2=l(l+n-1)-2$ where $l=2,3,...$. They satisfy
\begin{align}
\mathbb{T}_i^i=0 ;~~\bar{D}_j\mathbb{T}_i^j=0.
\label{eq19}
\end{align}

Henceforth, we shall consider $n > 2$. The analysis that follows can also be done for $n=2$ --- the only change being that we do not have the tensor spherical harmonics in that case.
The metric perturbations are decomposed as:
\begin{eqnarray}
{^{(i)}}h_{ab}=\sum_{\bf{k}_s}{^{(i)}}\!f_{ab\bf{k}_s}\mathbb{S}_{\bf{k}_s} ; ~~
{^{(i)}}h_{ai}=r\bigg(\sum_{\bf{k}_s}{^{(i)}}\!f_{a\bf{k}_s}\mathbb{S}_{\textbf{k}_si}+\sum_{\bf{k}_v}{^{(i)}}\!f_{a\bf{k}_v}^{(v)}\mathbb{V}_{\textbf{k}_vi}\bigg)\nonumber
\end{eqnarray}
\begin{align}
{^{(i)}}h_{ij}=&r^2\bigg(\sum_{\textbf{k}}{^{(i)}}\!H_{T\textbf{k}}\mathbb{T}_{\textbf{k}ij}+2\sum_{\textbf{k}_v}{^{(i)}}H^{(v)}_{T\textbf{k}_v}\mathbb{V}_{\textbf{k}_vij}\nonumber\\&+2\sum_{\textbf{k}_s}( {^{(i)}}\!H^{(s)}_{T\textbf{k}_s}\mathbb{S}_{\textbf{k}_sij}+{^{(i)}}H_{L\textbf{k}_s}\gamma_{ij}\mathbb{S}_{\textbf{k}_s})\bigg).
\label{eq20}
\end{align}
In the following sections, we drop the subscripts $\textbf{k}$, $\textbf{k}_v$ and $\textbf{k}_s$ from the metric perturbations to avoid cluttering the equations.\\\\
We now consider these perturbations in presence of a spherical Dirichlet wall of radius R. We need to fix the metric induced on a surface of radius $r=\text{R}$, requiring
\begin{align}
{^{(i)}}\!f_{tt}={^{(i)}}\!f_t={^{(i)}}\!H_T^{(s)}={^{(i)}}H_L={^{(i)}}\!f_t^{(v)}={^{(i)}}\!H_T^{(v)}={^{(i)}}\!H_T=0\big|_{r=\text{R}}
\label{eq74.11}
\end{align}
However, the metric components are also gauge dependent.
Under an infinitesimal gauge transformation $\bar{\delta}z^{\alpha}=\sum_{i}{^{(i)}}\zeta^{\alpha}$, metric perturbation ${^{(i)}}h_{\mu\nu}$ transforms as
\begin{align}
{^{(i)}}h_{\mu\nu}\rightarrow {^{(i)}}h_{\mu\nu}-\bar{\nabla}_{\mu}{^{(i)}}\zeta_{\nu}-\bar{\nabla}_{\nu}{^{(i)}}\zeta_{\mu}
\label{eq21},
\end{align}
i.e
\begin{align}
{^{(i)}}h_{ab}\rightarrow &{^{(i)}}h_{ab}-\bar{D}_a{^{(i)}}\zeta_b-\bar{D}_b{^{(i)}}\zeta_a,\nonumber\\
{^{(i)}}h_{ai}\rightarrow &{^{(i)}}h_{ai}-\bar{D}_i{^{(i)}}\zeta_a-r^2\bar{D}_a\left(  \frac{{^{(i)}}\zeta_i}{r^2}\right),\nonumber\\
{^{(i)}}h_{ij}\rightarrow & {^{(i)}}h_{ij}-\bar{D}_i{^{(i)}}\zeta_j-\bar{D}_j{^{(i)}}\zeta_i\nonumber\\&-2r\bar{D}^ar\:{^{(i)}}\zeta_a\gamma_{ij}.
\label{eq22}
\end{align}
Let ${^{(i)}}\zeta_a={^{(i)}}T_a\mathbb{S}$ and ${^{(i)}}\zeta_i=r\:{^{(i)}}\!L\mathbb{S}_i+r\:{^{(i)}}\!L^{(v)}\:\mathbb{V}_i$. Thus the gauge transformations for ${^{(i)}}f_{ab}$, ${^{(i)}}f_a$, ${^{(i)}}f_a^{(v)}$, ${^{(i)}}H_T^{(s)}$, ${^{(i)}}H_T^{(v)}$, ${^{(i)}}H_L$ and ${^{(i)}}H_T$ are
\begin{align}
{^{(i)}}f_{ab}&\rightarrow  {^{(i)}}f_{ab}-\bar{D}_a{^{(i)}}T_b-\bar{D}_b{^{(i)}}T_a,\\
{^{(i)}}f_a &\rightarrow {^{(i)}}f_a-r\bar{D}_a\left(  \frac{{^{(i)}}L}{r}\right)+\frac{k_s}{r}{^{(i)}}T_a,\\
{^{(i)}}H_L &\rightarrow {^{(i)}}H_L-\frac{k_s}{nr}{^{(i)}}L-\frac{\bar{D}^a r}{r}{^{(i)}}T_a,\\
{^{(i)}}H_T^{(s)}&\rightarrow {^{(i)}}H_T^{(s)}+\frac{k_s}{r}{^{(i)}}L,\\
{^{(i)}}f_a^{(v)}&\rightarrow{^{(i)}} f_a^{(v)}-r\bar{D}_a\left(\frac{{^{(i)}}L^{(v)}}{r}\right),\\
{^{(i)}}H_T^{(v)}&\rightarrow {^{(i)}}H_T^{(v)}+\frac{k_v}{r}{^{(i)}}L^{(v)},\\
{^{(i)}}H_T &\rightarrow {^{(i)}}H_T.
\label{eq23}
\end{align}\\\\
For all cases except $l_s=0,1$ and $l_v=1$ modes, one can define the following gauge invariant variables:

\begin{align}
{^{(i)}}\!Z_a={^{(i)}}\!f_a^{(v)}+\frac{r}{k_v}\bar{D}_a{^{(i)}}\!H_T^{(v)},
\label{eq23.1}
\end{align}
\begin{align}
{^{(i)}}F_{ab}={^{(i)}}\!f_{ab}+2\bar{D}_{(a}{^{(i)}}\!X_{b)};~~{^{(i)}}\!F={^{(i)}}H_L+\frac{{^{(i)}}\!H_T^{(s)}}{n}+\frac{1}{r}\bar{D}^ar{^{(i)}}\!X_a,
\label{eq23.2}
\end{align}
where ${^{(i)}}\!X_a$ is defined as
\begin{align}
{^{(i)}}X_a=\frac{r}{k_s}\left({^{(i)}}\!f_a+\frac{r}{k_s}\bar{D}_a{^{(i)}}\!H_T^{(s)}\right).
\label{eq23.3}
\end{align}
Since the expressions for $\triangle_L{^{(1)}}\!h_{\mu\nu}$ have already been given in terms of the gauge invariant variables in \cite{seto}, one can similarly get $\triangle_L{^{(i)}}\!h_{\mu\nu}$ in terms of the $i^{th}$ order gauge invariant variables ${^{(i)}}\!H_T$, ${^{(i)}}\!Z_a$, ${^{(i)}}\!F_{ab}$ and ${^{(i)}}\!F$ by replacing ${^{(1)}}\!h_{\mu\nu}$ by ${^{(i)}}h_{\mu\nu}$.
After we have solved for these variables we will use equations (\ref{eq23.1})-(\ref{eq23.3}) to obtain ${^{(i)}}\!h_{\mu\nu}$.  Before we do so, we can use the gauge freedom to put some terms to zero.
In subsequent sections, we will work in a gauge in which:
\begin{align}
{^{(i)}}\!f_t={^{(i)}}\!H_L={^{(i)}}\!H_T^{(s)}={^{(i)}}\!H_T^{(v)}=0.
\label{eq24}
\end{align}
(The above gauge choice is the same as used by \cite{soda} in case of vector perturbations and  \cite{marolf} in case of scalar perturbations.)

Our
strategy will be to first look at the equations for linearized modes.
The spectrum of linear perturbations is an important indicator of nonlinear stability.
If the spectrum is resonant, and if the higher order perturbation equations have the appropriate form that leads to
energy transfer to higher frequency modes, there
is likely to be an instability of the kind observed for $AdS$ spacetime \cite{bizon}. As we showed in earlier work \cite{ads} in
the context of the Einstein-Klein-Gordon system,
results from KAM theory indicate that when the spectrum is
not perfectly resonant but only approximately so, it is stable under arbitrarily small perturbations. However a
relatively small amplitude of perturbations may trigger instability that can be observed in numerical studies (as opposed to
arbitrarily small perturbations for a fully resonant spectrum). It is possible to quantify in number-theoretic terms how
close the spectrum is to being resonant. This is an indicator of the
magnitude of perturbation that may be required in order to trigger instability. A crucial ingredient in \cite{ads} was the structure of the
perturbation equations of the Einstein-Klein-Gordon system in weakly nonlinear perturbation theory. The
scalar field at third order obeys forced harmonic oscillator equations. The system can then be described by a Hamiltonian that is
a perturbation of an integrable Hamiltonian (linear harmonic oscillators) and it is possible to use Hamiltonian perturbation
theory to arrive at these conclusions. There are two questions of interest for our system:
what is the spectrum of the linearized perturbations, and
what is the structure of the higher order perturbation equations ? In subsequent sections, we will address both these questions.
By choosing appropriate master variables, we simplify the perturbation equations until the higher order equations for the master
variables resemble those of a forced harmonic oscillator (as in weakly nonlinear perturbation theory for the Einstein gravity-
scalar field system).
\section{Linearized  equations}
We will first have a look at the leading order equations in $\epsilon$.
\begin{align}
{^{(1)}}G_{\mu\nu}=\tilde{\triangle}_L {^{(1)}}h_{\mu\nu}=0.
\label{eq25}
\end{align}
Henceforth we drop the superscript "(1)" on ${^{(1)}}h_{\mu\nu}$.
Based on the methodology in \cite{seto}, linear perturbations of flat space in a cavity
with Dirichlet
boundary conditions at the cavity wall were studied in \cite{marolf}. In the following subsections, we revisit the linearized perturbations before moving to higher orders. In particular, we are interested in the spectrum of linearized perturbations as well as the linear order eigenfunctions as this is important to study the nonlinear
evolution of the perturbations. We use methods of Takahashi and Soda \cite{soda} for simplifying the equations for vector and scalar perturbations.
\subsection{Tensor perturbations}
In this case,                                                                                                                                                           \begin{align}
h_{ab}=0;~~ h_{ai}=0;~~ h_{ij}=\sum_{\bf{k}}r^2 H_{T\bf{k}}.\mathbb{T}_{\bf{k}ij}.
\label{eq28}
\end{align}
Upon substituting (\ref{eq28}) in the leading order in $\epsilon$ Einstein's equations, ${^{(1)}}G_{\mu\nu}$, we obtain the equation governing tensor type perturbations as given in \cite{seto}, wherein upon taking $H_{T\bf{k}}=\Phi_{T\textbf{k}}$ we get
\begin{align}
-r^2\ddot{\Phi}_T+r^2\Phi_T''+nr\Phi_T'-l(l+n-1)\Phi_T=0.
\label{eq29}
\end{align}
Equation (\ref{eq29}) can be put in the form
\begin{align}
\ddot{\Phi}_T+\hat{L}\Phi_T=0,
\label{eq30}
\end{align}
where
\begin{align}
\hat{L}=-\frac{1}{r^n}\partial_r(r^n\partial_r)+\frac{l(l+n-1)}{r^2}.
\label{eq30.1}
\end{align}
The solution to (\ref{eq30}) is given by
\begin{align}
\Phi_{T\textbf{k}}=\sum_{p=1}^{\infty}a_{p,\textbf{k}}\cos(\omega_{p,l}t+b_{p,\textbf{k}})e_{p,l}(r),
\label{eq31}
\end{align}
where $e_{p,l}(r)$ is given by
\begin{align}
e_{p,l}(r)=d_{p,l}\frac{J_{\nu}(\omega_{p,l} r)}{r^{(n-1)/2}};~~\nu=l+\frac{(n-1)}{2},
\label{eq32}
\end{align}
 Constants $a_{p,\textbf{k}}$ and $b_{p,\textbf{k}}$ are determined from initial conditions and $d_{p,l}$ is the normalization constant given by $\frac{\sqrt{2}}{\text{R}J_{\nu}'(\omega_{p,l}\text{R})}$.
The eigenfrequencies $\omega_{p,l}$ are discrete and associated with a mode number $p$ for each $l$. They are determined from the Dirichlet boundary condition: $\Phi_T=0$ at $r=\text{R}$.
\begin{align}
\Rightarrow \omega_{p,l} =\frac{j_{\nu,p}}{\text{R}},
\label{eq33}
\end{align}
where $j_{\nu,p}$ is the $p^{th}$ zero of a Bessel function of order $\nu$. For large values of $p$, eigenfrequencies approach $(p+\nu/2-1/4)\frac{\pi}{\text{R}}$, therefore
it is an asymptotically resonant spectrum.

The eigenfunctions $e_{p,l}$ form a complete set and are orthonormal in the space of functions $\hat{L}^2([0,\text{R}], r^{n}dr)$. The inner product $\left<f,g\right>_T$ is defined as $\int_0^{\text{R}}f(r)g(r)r^ndr$.

\subsection{Vector perturbations}
The following equations govern the linear vector perturbations
\cite{seto}:
\begin{align}
\bar{D}_a(r^{n-1}Z^a)=0,
\label{eq45l}
\end{align}
\begin{align}
-\frac{1}{r^{n}}\bar{D}^b\left[r^{n+2}\left[\bar{D}_b\left(\frac{Z_a}{r}\right)-\bar{D}_a\left(\frac{Z_b}{r}\right)\right]\right]+\frac{\left(k^2_v-(n-1)\right)}{r}Z_a=0.
\label{eq46l}
\end{align}
These, when expanded yield three equations. Of them, two
are independent. The third equation can be obtained from the first two
through a suitable combination. Here we will use only two of them.
One is obtained by expanding equation (\ref{eq45l}):
\begin{align}
\dot{Z}_t=(n-1)\frac{Z_r}{r}+Z_r'.
\label{eq47l}
\end{align}
The other is obtained by making the substitution $a=r$ in (\ref{eq46l}):
\begin{align}
-\partial_t^2 Z_r
+r\partial_t\partial_r\left(\frac{Z_t}{r}\right)-\frac{(k_v^2-(n-1))}{r^2}Z_r=0.
\label{eq48l}
\end{align}
Next we substitute the expression for $\dot{Z}_t$ from (\ref{eq47l}) in
(\ref{eq48l}), so that we get a second order equation in $Z_{r}$:
\begin{align}
-\ddot{Z_r}+Z_r''+\frac{(n-2)}{r}Z_r'-&\frac{(l_v(l_v+n-1)+(n-2))}{r^2}Z_r=0.
\label{eq49l}
\end{align}
We rewrite $Z_{r\textbf{k}_v}$ as
\begin{align}
Z_{r\textbf{k}_v}=r\Phi_{v\textbf{k}_v}.
\label{eqe49.1l}
\end{align}
Hence upon writing (\ref{eq49l}) in terms of the new variable
$\Phi_{v\textbf{k}_v}$, we obtain the following master equation,
\begin{align}
\ddot{\Phi}_v+\hat{L}_v\Phi_v=0,
\label{eq50l}
\end{align}
where $\hat{L}_v$ is defined as:
\begin{align}
\hat{L}_v=-\frac{1}{r^n}\partial_r(r^n\partial_r)+\frac{l_v(l_v+n-1)}{r^2}.
\label{eq51l}
\end{align}
Let $\Phi_{v\textbf{k}_v}=\cos(\omega t+b)\phi_{v\textbf{k}_v}(r)$, so that equation
(\ref{eq50l}) could be rewriiten as:
\begin{align}
\hat{L}_v\phi_v=\omega^2\phi_v.
\label{eq52l}
\end{align}
Demanding Dirichlet boundary condition is equivalent to imposing $Z_t=0$ at
$r=\text{R}$. In order to impose this condition, we will write (\ref{eq47l}) in terms of $\Phi_{v\textbf{k}_v}$ to get:
\begin{align}
\dot{Z}_t=r\Phi_v'+n\Phi_v.
\label{eqe52.1l}
\end{align}
Then using the ansatz $\Phi_{v\textbf{k}_v}= \cos(\omega t+b)\phi_{v\textbf{k}_v}(r)$ and then integrating w.r.t. $t$, we get,
\begin{align}
Z_t=\frac{1}{\omega}\{r\phi_v'+n\phi_v\} \sin(\omega t+b).
\label{eqe52.2l}
\end{align}
Any $r-$dependent integration constant in the above expression is put to zero. Hence, if Dirichlet condition on $Z_{t}$ needs to be satisfied at all times, we require

\begin{align}
r\phi_v'+n\phi_v=0 \big|_{r=\text{R}}.
\label{eq54.1l}
\end{align}
The linear stability of the vector modes has been shown in
\cite{marolf} so we will not repeat the argument here.
 The eigenfrequencies are discrete and hence can be associated with a mode number $p$ for each $l_v$. Hence, we find that
\begin{align}
\phi_{v\textbf{k}_v}=e^{(v)}_{p,l_v}=d^{(v)}_{p,l_v}\frac{
J_{\nu_v}(\omega_{p,l_v} r) }{r^{(n-1)/2}};~~\nu_v=l_v+\frac{(n-1)}{2},
\label{eq53.1l}
\end{align} where  $d^{(v)}_{p,l_v}$ is the normalization constant given by
\begin{align}
d^{(v)}_{p,l_v}=\frac{\sqrt{2}\omega_{p,l_v}}{ J_{\nu_v}(\omega_{p,l_v}\text{R})}\Big[ (n+1)^2/4+(\omega_{p,l_v}\text{R})^2-\nu_v^2\Big]^{-1/2}.
\end{align}

 The
eigenfunctions $\phi_{v\textbf{k}_v}=e^{(v)}_{p,l_v}(r)$ are complete and
form an orthonormal basis in the space of functions
$\hat{L}^2_v([0,\text{R}],r^ndr)$. Therefore, the general solution to (\ref{eq50l}) is given by
 \begin{align}
\Phi_{v\textbf{k}_v}=\sum_{p=1}^{\infty}a^{(v)}_{p,\textbf{k}_v}\cos(\omega_{p,l_v}
t+b_{p,\textbf{k}_v})e^{(v)}_{p,l_v}(r),
\label{eq53l}
\end{align}
where $a_{p,\textbf{k}_v}^{(v)}$ and $b_{p,\textbf{k}_v}$ are determined
from initial conditions and $\omega_{p,l_v}$ satisfies (\ref{eq54.1l}).
 The inner product $\left<f,g\right>_v$ is
defined as $\int_0^{\text{R}} f(r)g(r)r^n dr$.
Upon substituting for $\phi_v$ from (\ref{eq53.1l}) in (\ref{eq54.1l}) we
obtain
\begin{align}
\omega rJ_{\nu_v}'(\omega r)+\frac{(n+1)}{2}J_{\nu_v}(\omega r)=0
\big|_{r=\text{R}}.
\label{57l}
\end{align}

Now we will look at the asymptotic nature of the frequencies associated with vector modes by
considering the large argument expansion of the Bessel functions, which is given by

\begin{align}
J_{\nu_v}(z)\sim\sqrt{\frac{2}{z\pi}}\cos\left(z-\frac{\nu_v\pi}{2}-\frac{\pi}{4}\right) \text{
as  } z\rightarrow \infty.
\label{eq53.3l}
\end{align}
This tells us that for large modes
\begin{align}
\tan\left(z-\frac{\nu_v\pi}{2}-\frac{\pi}{4}\right)=\frac{(n+1)/2}{z},
\label{eq53.4l}
\end{align}
where $z=\omega \text{R}$. It can be seen that the frequencies tend to
$\left(p+\frac{\nu_v}{2}-\frac{3}{4}\right)\frac{\pi}{\text{R}}$.
\subsection{Scalar perturbations}
Using the Ishibashi-Kodama-Seto formalism \cite{seto}, we get the scalar perturbation equations. In order to obtain the master equation, we will use the method followed by Takahashi and Soda \cite{soda}.
Scalar perturbations satisfy the following identity
\begin{align}
\left[2(n-2)F+F^c_c\right]=0,
\label{eq55l}
\end{align}
which is obtained from (traceless part of) the ${^{(1)}}G_{ij}=0$ equation.
From ${^{(1)}}\!G_{rt}=0$ one gets
\begin{align}
\frac{n}{r}\dot{F}_{rr}+\frac{k_s^2}{r^2}F_{rt}-2n\dot{F}'-\frac{2n}{r}\dot{F}=0.
\label{eq56l}
\end{align}
Similar to \cite{soda}, we choose
\begin{align}
F_{rt}=2r(\dot{\Phi}_s+\dot{F}),
\label{eq58l}
\end{align}
where $\Phi_s$ is our master variable.
This helps us to integrate (\ref{eq56l}) with respect to $t$ and get an expression for $F_{rr}$ in terms of $F$ and $\Phi_s$ which is:
\begin{align}
F_{rr}=2rF'+2F-\frac{2k_S^2}{n}F-\frac{2k^2_s}{n}\Phi_s.
\label{eq59l}
\end{align}
The extra integration constant, which would be a function of $r$, is absorbed in the definition of $\Phi_s$.\\
From ${^{(1)}}\!G_{tt}=0$ one gets
\begin{align}
-2nF''+\frac{n}{r}F_{rr}'+\Big(\frac{k_s^2}{r^2}&+\frac{n(n-1)}{r^2}\Big)F_{rr}-\frac{2n(n+1)}{r}F'\nonumber\\&+\left(\frac{2k_s^2(n-1)}{r^2}-\frac{2n(n-1)}{r^2}\right)F=0.
\label{eq60l}
\end{align}
Substituting the expression for $F_{rr}$ from (\ref{eq59l}) into (\ref{eq60l}) gives us an expression for $F$ solely in terms of $\Phi_s$ and its derivative:
\begin{align}
F=-\frac{n}{k_s^2-n}\left[r\Phi_s'+\left(\frac{k_s^2}{n}+n-1\right)\Phi_s\right].
\label{eq61l}
\end{align}
In the scalar component of ${^{(1)}}\!G_{ir}=0$ we will substitute for $F_c^c$ from (\ref{eq55l}). This gives:
\begin{align}
-\frac{(n-2)}{r}F_{rr}+\dot{F}_{rt}-F_{rr}'+2F'+\frac{2(n-2)}{r}F=0.
\label{eq63l}
\end{align}
Next, by using ${^{(1)}}G_{rr}=0$ and (\ref{eq55l}), one gets
\begin{align}
-2n\ddot{F}+\frac{2n}{r}F'-\frac{2k_s^2}{r^2}F&+\frac{2n(n-1)}{r^2}F-\frac{n}{r}F_{rr}'\nonumber\\&+\left(\frac{k_s^2}{r^2}-\frac{n(n-1)}{r^2}\right)F_{rr}+\frac{2n}{r}\dot{F_{rt}}=0.
\label{eq66l}
\end{align}
We eliminate $F_{rr}'$ from (\ref{eq66l}) by using (\ref{eq63l}). This gives
\begin{align}
-2n\ddot{F}+\frac{(k_s^2-n)}{r^2}F_{rr}-\frac{2(k_s^2-n)}{r^2}F+\frac{n}{r}\dot{F}_{rt}=0.
\label{eq68l}
\end{align}
Next, using the expressions for $F_{rt}$, $F_{rr}$ and $F$  from (\ref{eq58l}), (\ref{eq59l}) and (\ref{eq61l}) in (\ref{eq68l}) leads to the second order master equation for $\Phi_{s\textbf{k}_s}:$
\begin{align}
\ddot{\Phi}_s-\Phi_s''-\frac{n}{r}\Phi_s'+\frac{l_s(l_s+n-1)}{r^2}\Phi_s=0.
\label{eq70l}
\end{align}
We can then rewrite (\ref{eq70l}) as
\begin{align}
\ddot{\Phi}_s+\hat{L}_s\Phi_s=0,
\label{eq72l}
\end{align}
where $\hat{L}_s=-\frac{1}{r^n}\partial_r(r^n\partial_r)+\frac{l_s(l_s+n-1)}{r^2}$.\\\\ We substitute the ansatz  $\Phi_s=\cos(\omega t+b)\phi_s$ in (\ref{eq72l}) and get,
\begin{align}
\hat{L}_s\phi_s=\omega^2\phi_s
\label{eq74l}
\end{align}

The eigenfrequencies $\omega$ must satisfy the mixed frequency dependent boundary condition obtained by requiring $f_{tt}$ (or equivalently $F_{tt}$) to vanish at the boundary $r=\text{R}$. This is given by:
\begin{align}
(n-1)r\phi_s'+\left(-\omega^2r^2+\frac{(n-1)}{n}(k_s^2+n(n-1))\right)\phi_s=0\Big|_{r=\text{R}}.
\label{eq73.2l}
\end{align}
In \cite{marolf}, the stability under scalar perturbations has been demonstrated. The eigenfrequencies are discrete and can be associated with the mode number, say $p$ for each $l_s$.
Numerically, it can be seen that the spectrum is asymptotically resonant and high frequencies approach
$\left(p+\frac{\nu_s}{2}-\frac{5}{4}\right)\frac{\pi}{\text{R}}$.

The eigensolutions of equation (\ref{eq74l}) are
\begin{align}
\phi_{s\textbf{k}_s}=e^{(s)}_{p,l_s}(r)=d^{(s)}_{p,l_s}\frac{J_{\nu_s}(\omega_{p,l_s}r)}{r^{(n-1)/2}} ;~~\nu_s=l_s+\frac{(n-1)}{2}.
\label{eq73.1l}
\end{align}
where the constant $d^{(s)}_{p,l_s}$ is given by
\begin{align}
d_{p,l_s}^{(s)}=\left[\int_0^{\text{R}}|J_{\nu_s}(\omega_{p,l_s}r)|^2rdr+\frac{\text{R}^{2}}{(n-1)}
|J_{\nu_s}(\omega_{p,l_s}\text{R})|^2\right]^{-1/2}
\label{add2}
\end{align}

Although the eigenfunctions $e_{p,l_s}^{(s)}$ are not orthogonal, they satisfy the modified orthogonality relation,
$<e_{p,l_s}^{(s)},e_{q,l_s}^{(s)}>_s$, given by
\begin{align}
<e_{p,l_s}^{(s)},e_{q,l_s}^{(s)}>_s=\int_0^{\text{R}}e_{p,l_s}^{(s)}e_{q,l_s}^{(s)}r^ndr+\frac{\text{R}^{n+1}}{(n-1)}e_{p,l_s}^{(s)}(\text{R})e_{q,l_s}^{(s)}(\text{R})=
\begin{cases}
0 &\text{ for $p\neq q$}\\
1  &\text{for $p=q$}
\end{cases}
\label{add1}
\end{align}

We can write the general solution to (\ref{eq74l}) in terms of a series expansion of the discrete modes.
We use the work of Zecca \cite{zecca} which deals with the
Bessel equation in a finite interval with singularity at one end and a
eigenvalue dependent boundary condition, similar to ours, at the regular end
point. He shows that the general solution can be expanded in a series of Bessel functions within this finite interval.
Therefore, the solution to equation (\ref{eq72l}) is
\begin{align}
\Phi_{s\textbf{k}_s}=\sum_{p=0}^{\infty}a_{p,\textbf{k}_s}^{(s)}\cos(\omega_{p,l_s} t+b_{p,\textbf{k}_s})e^{(s)}_{p,l_s}(r),
\label{eq73l}
\end{align}
where $a^{(s)}_{p,\textbf{k}_s}$ and $b_{p,\textbf{k}_s}$ are constants set by initial conditions.
An expansion theorem in his paper then implies that a function $f(r)$ in $C^{1}[0,1]$
with square integrable second derivative, and which satisfies the same
boundary conditions as $J_{\nu_s}(\omega_{p,l_s}r)$ (or $ e^{(s)}_{p,l_s}$)
can be expanded in a series of these Bessel functions (or eigenfunctions
$e_{p,l_s}^{(s)}$). Hence we can write $f(r)$ as
\begin{align}
f=\sum_{p}\sigma_p e_{p,l_s}^{(s)}
\label{add3}
\end{align}
where
\begin{align}
\sigma_p=<f,e_{p,l_s}^{(s)}>_s=\int_0^{\text{R}}fr^ne_{p,l_s}^{(s)}dr+\frac{\text{R}^{n+1}}{(n-1)}f(\text{R})e_{p,l_s}^{(s)}(\text{R})
\label{add4}
\end{align}
We will be using the above results when dealing with higher order perturbations.
\section{Higher order equations}
The higher order perturbed equations have to be solved for ${^{(i)}}\!h_{\mu\nu}$ given a source which is composed of $1,..  ,(i-1)^{th}$ order  metric perturbations. For eg.,
the second order perturbed equations, ${^{(2)}}G_{\mu\nu}=0$ become
\begin{align}
{^{(2)}}G_{\mu\nu}=\tilde{\triangle}_L {^{(2)}}h_{\mu\nu}-\frac{1}{2}{^{(2)}}S_{\mu\nu}=0,
\label{eq35}
\end{align}
where ${^{(2)}}\!S_{\mu\nu}$ is
\begin{align}
{^{(2)}}\!S_{\mu\nu}=&-\frac{1}{2}\bar{\nabla}_{\lambda}h(-\bar{\nabla}^{\lambda}h_{\mu\nu}+\bar{\nabla}_{\nu}h_{\mu}^{\lambda}+\bar{\nabla}_{\mu}h^{\lambda}_{\nu})-\frac{1}{2}\bar{\nabla}_{\nu}h^{\lambda\sigma}\bar{\nabla}_{\mu}h_{\sigma\lambda}-\bar{\nabla}^{\sigma}h_{\mu\lambda}\bar{\nabla}_{\sigma}h_{\nu}^{\lambda}\nonumber\\&+\bar{\nabla}_{\lambda}h_{\mu}^{\sigma}\bar{\nabla}_{\sigma}h_{\nu}^{\lambda}+\bar{\nabla}_{\lambda}h^{\lambda\sigma}(-\bar{\nabla}_{\sigma}h_{\mu\nu}+\bar{\nabla}_{\nu}h_{\mu\sigma}+\bar{\nabla}_{\mu}h_{\sigma\nu})\nonumber\\&+h^{\lambda\sigma}(-\bar{\nabla}_{\lambda}\bar{\nabla}_{\sigma}h_{\mu\nu}-\bar{\nabla}_{\mu}\bar{\nabla}_{\nu}h_{\lambda\sigma}+\bar{\nabla}_{\lambda}\bar{\nabla}_{\mu}h_{\sigma\nu}+\bar{\nabla}_{\sigma}\bar{\nabla}_{\nu}h_{\mu\lambda})\nonumber\\&+\frac{1}{2}\eta_{\mu\nu}\bigg(-\frac{1}{2}\bar{\nabla}_{\alpha}h\bar{\nabla}^{\alpha}h+2\bar{\nabla}_{\alpha}h^{\alpha\beta}\bar{\nabla}_{\beta}h+\frac{3}{2}\bar{\nabla}_{\alpha}h_{\beta\sigma}\bar{\nabla}^{\alpha}h^{\beta\sigma}\nonumber\\&-\bar{\nabla}_{\alpha}h_{\beta}^{\sigma}\bar{\nabla}_{\sigma}h^{\alpha\beta}-2\bar{\nabla}_{\alpha}h^{\alpha}_{\sigma}\bar{\nabla}_{\beta}h^{\beta\sigma}+h^{\lambda\sigma}\bar{\nabla}_{\lambda}\bar{\nabla}_{\sigma}h+h^{\lambda\sigma}\bar{\nabla}^{\alpha}\bar{\nabla}_{\alpha} h_{\lambda\sigma}\nonumber\\& -2h^{\alpha\beta}\bar{\nabla}_{\alpha}\bar{\nabla}_{\sigma}h^{\sigma}_{\beta}  \bigg).
\label{eq36}
\end{align}

The details of the calculation are given in appendix A. Now we look at a general $i^{th}$ order equation where $i\geq 2$. \\

Equation (\ref{eq6.1}) can be written in terms of $\bar{D}_a$ and $\bar{D}_i$ operators. Using the expansion of $\triangle_L h_{\mu\nu}$  given in the appendix of \cite{seto} we obtain the perturbed equations, ${^{(i)}}\!G_{\mu\nu}=0$  in terms of ${^{(i)}}\!H_T$, ${^{(i)}}\!F$, ${^{(i)}}\!F_{ab}$ and ${^{(i)}}\!Z_a$.
For $\mu=i,\nu=j$ one obtains
\begin{align}
&\sum_{\textbf{k}}  \Bigg[-r^2\bar{D}^a\bar{D}_a{^{(i)}}\!H_T-nr\bar{D}^ar\bar{D}_a{^{(i)}}\!H_T+(k^2+2K){^{(i)}}\!H_T\Bigg]_{\textbf{k}}\mathbb{T}_{\textbf{k}ij}\nonumber\\&+\sum_{\textbf{k}_v} \Bigg[-\frac{2k_v}{r^{n-2}}\bar{D}_a(r^{n-1}{^{(i)}}\!Z^a)\Bigg]_{\textbf{k}_v}\mathbb{V}_{\textbf{k}_vij}+\sum_{\textbf{k}_s}\Bigg[-k_s^2[2(n-
2){^{(i)}}\!F+{^{(i)}}\!F_c^c]\Bigg]_{\textbf{k}_s}\mathbb{S}_{\textbf{k}_sij}\nonumber\\&={^{(i)}}\!S_{ij}-\sum_{\textbf{k}_s}[Q_4]_{\textbf{k}_s}\gamma_{ij}\mathbb{S}_{\textbf{k}_s}.
\label{eqa10}
\end{align}
 We do not write the explicit form of $[Q_4]$  as it is not required in our
calculations and it does not contribute when we finally project to individual tensor components of each type.\\
Similarly from ${^{(i)}}G_{ai}=0$ equation one gets
\begin{align}
&\sum_{\textbf{k}_v}\Bigg[-\frac{1}{r^n}\bar{D}^b\left\{ r^{n+2}\left[\bar{D}_b\left(\frac{{^{(i)}}\!Z_a}{r}\right)-\bar{D}_a\left(\frac{{^{(i)}}\!Z_b}{r}\right)\right]\right\}+\frac{k^2_v-(n-1)K}{r}{^{(i)}}\!Z_a\Bigg]_{\textbf{k}_v}\mathbb{V}_{\textbf{k}_vi}\nonumber\\&+\sum_{\textbf{k}_s}\Bigg[-k_s\Big(\frac{1}{r^{n-2}}\bar{D}_b(r^{n-2}{^{(i)}}\!F^b_a)-r\bar{D}_a\bigg(\frac{1}{r}{^{(i)}}\!F^b_b\bigg)-2(n-1)\bar{D}_a{^{(i)}}\!F\Big)
\Bigg]_{\textbf{k}_s}\mathbb{S}_{\textbf{k}_si}={^{(i)}}\!S_{ai}.
\label{eqa11}
\end{align}
In order to decompose the various sectors we use the fact that
\begin{align}
\int \mathbb{T}^{ij}\mathbb{V}_{ij}d^n\Omega=\int \mathbb{T}^{ij}\mathbb{S}_{ij}d^n\Omega=\int \mathbb{V}^{ij}\mathbb{S}_{ij}d^n\Omega=\int \mathbb{V}^i\mathbb{S}_id^n\Omega=0.
\label{eq36.4}
\end{align}
We obtain the tensor equation from (\ref{eqa10}) which is
\begin{align}
-r^2\bar{D}^a\bar{D}_a{^{(i)}}\!H_T-nr\bar{D}^ar\bar{D}_a{^{(i)}}\!H_T+(k^2+2K){^{(i)}}\!H_T=\int \mathbb{T}^{ij}_{\bf{k}}\:{^{(i)}}\!S_{ij}d^n\Omega.
\label{eq37}
\end{align}
Similarly using (\ref{eq36.4}) we obtain the two vector equations from (\ref{eqa10}) and (\ref{eqa11})
\begin{align}
-\frac{1}{r^n}\bar{D}^b\left\{ r^{n+2}\left[\bar{D}_b\left(\frac{{^{(i)}}\!Z_a}{r}\right)-\bar{D}_a\left(\frac{{^{(i)}}\!Z_b}{r}\right)\right]\right\}&+\frac{k^2_v-(n-1)K}{r}{^{(i)}}\!Z_a\nonumber\\&=\int \mathbb{V}^i_{\bf{k}_v}{^{(i)}}\!S_{ai} d^n\Omega,
\label{eq38}
\end{align}
\begin{align}
-\frac{2k_v}{r^{n-2}}\bar{D}_a(r^{n-1}{^{(i)}}\!Z^a)=\int \mathbb{V}^{ij}_{\bf{k}_v}{^{(i)}}\!S_{ij} d^n\Omega.
\label{eq39}
\end{align}

For the scalar case we will use the following three equations, which are
\begin{align}
&-\bar{D}^c\bar{D}_c {^{(i)}}\!F_{ab}+\bar{D}_a\bar{D}_c {^{(i)}}\!F^c_b+\bar{D}_b\bar{D}_c {^{(i)}}\!F^c_a+n\frac{\bar{D}^cr}{r}(-\bar{D}_c {^{(i)}}\!F_{ab}+\bar{D}_a {^{(i)}}\!F_{cb}+\bar{D}_b {^{(i)}}\!F_{ca})\nonumber\\&+\frac{k_s^2}{r^2}{^{(i)}}\!F_{ab}-\bar{D}_a\bar{D}_b {^{(i)}}\!F^c_c-2n\left(   \bar{D}_a\bar{D}_b {^{(i)}}\!F+\frac{1}{r}\bar{D}_a r\bar{D}_b{^{(i)}}\!F+\frac{1}{r}\bar{D}_b r\bar{D}_a{^{(i)}}\!F\right)\nonumber\\&-\Bigg( \bar{D}_c\bar{D}_d {^{(i)}}\!F^{cd}+\frac{2n}{r}\bar{D}^cr\bar{D}^d {^{(i)}}\!F_{cd}+\frac{n(n-1)}{r^2}\bar{D}^cr\bar{D}^d r {^{(i)}}\!F_{cd} -2n\bar{D}^c\bar{D}_c {^{(i)}}\!F\nonumber\\&-\frac{2n(n+1)}{r}\bar{D}^cr\bar{D}_c{^{(i)}}\!F+2(n-1)\frac{(k^2_s-nK)}{r^2}{^{(i)}}\!F-\bar{D}^c\bar{D}_c {^{(i)}}\!F^d_d\nonumber\\&-\frac{n}{r}\bar{D}^c r \bar{D}_c {^{(i)}}\!F^d_d+\frac{k^2_s}{r^2}{^{(i)}}\!F^d_d\Bigg)\eta_{ab}=\int \mathbb{S}_{\bf{k}_s}{^{(i)}}\!S_{ab} d^n\Omega,
\label{eq40}
\end{align}
\begin{align}
-k_s\Big(\frac{1}{r^{n-2}}\bar{D}_b(r^{n-2}{^{(i)}}\!F^b_a)-r\bar{D}_a\left(\frac{1}{r}{^{(i)}}\!F^b_b\right)&-2(n-1)\bar{D}_a {^{(i)}}\!F\Big)\nonumber
\\&=\int \mathbb{S}_{\bf{k}_s}^i {^{(i)}}\!S_{ai} d^n\Omega,
\label{eq41}
\end{align}
\begin{align}
-k^2_s[2(n-2){^{(i)}}\!F+{^{(i)}}\!F^c_c]=\int \mathbb{S}^{ij}_{\bf{k}_s}{^{(i)}}\!S_{ij} d^n\Omega.
\label{eq42}
\end{align}

\subsection{Tensor perturbations at higher orders}
Let ${^{(i)}}\!H_{T\textbf{k}}={^{(i)}}\!\Phi_{T\textbf{k}}$. Thus (\ref{eq37}) becomes
\begin{align}
{^{(i)}}\ddot{\Phi}_{T\textbf{k}}+\hat{L}{^{(i)}}\Phi_{T\textbf{k}}=\frac{1}{r^2}\int \mathbb{T}^{ij}_{\textbf{k}}\:{^{(i)}}\!S_{ij}d^n\Omega.
\label{eq44.1}
\end{align}
Condition (\ref{eq74.11}) implies that at each order $\Phi_{T\textbf{k}}$ should vanish at the boundary $r=\text{R}$.
The set of eigenfunctions $e_{p,l}(r)$ are complete and also satisfy condition (\ref{eq74.11}). Hence, one can expand $\Phi_{T\textbf{k}}$ as:
\begin{align}
{^{(i)}}\Phi_{T\textbf{k}}=\sum_{p=1}^{\infty}{^{(i)}}\!c_{p,\textbf{k}}(t)e_{p,l}(r).
\label{eqe44.1}
\end{align}
Hence (\ref{eq44.1}) can be written as
\begin{align}
{^{(i)}}\!\ddot{c}_{p,\textbf{k}}(t)+\omega^2_{p,l}{^{(i)}}\!c_{p,\textbf{k}}(t)=\left< \frac{1}{r^2}\int \mathbb{T}^{ij}_{\textbf{k}}\:{^{(i)}}\!S_{ij}, e_{p,l}(r)\right>_T.
\label{eq44.2}
\end{align}
\subsection{Vector perturbations at higher  orders}
Before we proceed, we will define the following two quantities:
\begin{align}
{^{(i)}}\!V_{s1}=\int \mathbb{V}^{ij}_{\textbf{k}_v}{^{(i)}}\!S_{ij} d^n\Omega,
\label{eq47.1}
\end{align}
\begin{align}
{^{(i)}}\!V_{s2}=\int \mathbb{V}^i_{\textbf{k}_v}{^{(i)}}S_{ri}d^n\Omega.
\label{eq47.2}
\end{align}
We firstly expand equation (\ref{eq39}) and obtain the following:
\begin{align}
{^{(i)}}\!\dot{Z}_t=(n-1)\frac{{^{(i)}}\!Z_r}{r}+{^{(i)}}\!Z_r'+\frac{1}{2k_vr}{^{(i)}}\!V_{s1}.
\label{eq48}
\end{align}
(Note that in the preceding equation as well as in the equations which follow, we drop the subscript $\textbf{k}_v$ in $Z_{a\textbf{k}_v}$ for convenience.) Next, by making the substitution $a=r$ in (\ref{eq38}), we obtain:
\begin{align}
- {^{(i)}}\!\ddot{Z}_r + r\partial_t\partial_r\left(\frac{{^{(i)}}\!Z_t}{r}\right)-\frac{(k_v^2-(n-1))}{r^2}{^{(i)}}\!Z_r=-\frac{{^{(i)}}\!V_{s2}}{r}.
\label{eq49}
\end{align}
Now we substitute the expression for ${^{(i)}}\!\dot{Z}_t$ from (\ref{eq48}) in (\ref{eq49}) to get:
\begin{align}
-{^{(i)}}\!\ddot{Z_r}+{^{(i)}}\!Z_r''+\frac{(n-2)}{r}{^{(i)}}\!Z_r'-&\frac{(l_v(l_v+n-1)+(n-2))}{r^2}{^{(i)}}\!Z_r\nonumber\\&=-\left[\frac{{^{(i)}}\!V_{s2}}{r}+r\left( \frac{{^{(i)}}\!V_{s1}}{2k_v r^2}\right)'\right]
\label{eq50}.
\end{align}
Now we rewrite ${^{(i)}}\!Z_{r\textbf{k}_v}$ in terms of a new variable ${^{(i)}}\!\Phi_{s\textbf{k}_s}$ as
\begin{align}
{^{(i)}}\!Z_{r\textbf{k}_v}=r\left({^{(i)}}\!\Phi_{v\textbf{k}_v}-\frac{1}{2k_vr^n}\int {^{(i)}}\!V_{s1}r^{(n-2)}dr\right ).
\label{eq51}
\end{align}
The above definition is crucial, since it enables us to expand the higher order perturbations ${^{(i)}}\!\Phi_{v\textbf{k}_v}$ in terms of the complete set of eigenfunctions $e_{p,l_v}^{(v)}$. This will be made clear in a while. Substitution of (\ref{eq51}) in (\ref{eq50}) leads to the following equation in terms of the variable ${^{(i)}}\!\Phi_{v\textbf{k}_v}$:
\begin{align}
{^{(i)}}\!\ddot{\Phi}_{v\textbf{k}_v}+\hat{L}_v{^{(i)}}\!\Phi_{v\textbf{k}_v}=\frac{1}{r}\left[\frac{{^{(i)}}\!V_{s2}}{r}+r\left(\frac{{^{(i)}}\!V_{s1}}{2k_v r^2}\right)'\right]&+\frac{\int{^{(i)}}\!\ddot{V}_{s1}r^{(n-2)}dr}{2k_vr^n}\nonumber\\&+\hat{L}_v\left[\frac{\int {^{(i)}}\!V_{s1}r^{(n-2)}dr}{2k_vr^n}\right].
\label{eq52}
\end{align}
Further simplification of the above equation can be done by expanding ${^{(i)}}\!\Phi_{v\textbf{k}_v}$ in the basis of a complete set of functions, (which also satisfy the appropriate boundary condition) say ${^{(i)}}\!\phi_{v\textbf{k}_v}$ as follows:
\begin{align}
{^{(i)}}\!\Phi_{v\textbf{k}_v}=\sum_{p=1}^{\infty}{^{(i)}}\!c_{p,\textbf{k}_v}^{(v)}(t){^{(i)}}\!\phi_{v\textbf{k}_v}(r).
\label{eq53}
\end{align}
We substitute for ${^{(i)}}\!Z_r$  from (\ref{eq51}) in (\ref{eq48}) and use the expansion (\ref{eq53}). The expression for ${^{(i)}}\!Z_{t\textbf{k}_v}$ then becomes
\begin{align}
{^{(i)}}\!Z_{t\textbf{k}_v}=\sum_{p=1}^{\infty} \int {^{(i)}}\!c_{p,\textbf{k}_v}(t) dt\{r{^{(i)}}\!\phi_{v\textbf{k}_v}'+n{^{(i)}}\!\phi_{v\textbf{k}_v}\}
\label{eqe52.1}
\end{align} In the above equation any $r-$dependent integration constant is put to zero. Then
we apply Dirichlet boundary condition that requires ${^{(i)}}\!Z_t$ to vanish at $r=\text{R}$ for all times. This means
\begin{align}
r{^{(i)}}\!\phi_v'+n{^{(i)}}\!\phi_v=0\big|_{r=\text{R}}.
\label{eq53.1}
\end{align}
Note that the ansatz (\ref{eq51}) has specifically been chosen so that the boundary condition takes the above form.
Since the eigenfunctions corresponding to the linear perturbation $\phi_v=e_{p,l_v}^{(v)}(r)$ (which form a complete set) also satisfy (\ref{eq53.1}), we can choose to expand metric perturbations in this basis. i.e., choose
${^{(i)}}\phi_v=\phi_v$. Taking the projection of (\ref{eq52}) on $e_{p,l_v}^{(v)}$ one gets a forced harmonic oscillator equation of the form
\begin{align}
{^{(i)}}\!\ddot{c}_{p,\textbf{k}_v}^{(v)}+\omega_{p,l_v}^2{^{(i)}}\!c_{p,\textbf{k}_v}^{(v)}=\Bigg<\Bigg[\frac{{^{(i)}}\!V_{s2}}{r^2}+\Big(\frac{{^{(i)}}\!V_{s1}}{2k_vr^2}\Big)'\Bigg]&+\frac{\int{^{(i)}}\!\ddot{V}_{s1}r^{(n-2)}dr}{2k_vr^n}\nonumber\\&+\hat{L}_v\Bigg[\frac{\int {^{(i)}}\!V_{s1}r^{(n-2)}dr}{2k_vr^n} \Bigg], e_{p,l_v}^{(v)}\Bigg>_v.
\label{eq54}
\end{align}
\subsection{Scalar perturbations at higher order}
Before we consider the scalar equations we will define the following quantities:
\begin{align}
{^{(i)}}\!S_{s0}=\int \mathbb{S}^{ij}_{\bf{k}_s}{^{(i)}}\!S_{ij}d^n\Omega,
\label{eq54.1}
\end{align}
\begin{align}
{^{(i)}}\!S_{s1}=\int \mathbb{S}_{\bf{k}_s}{^{(i)}}\!S_{rt} d^n\Omega,
\label{eq54.2}
\end{align}
\begin{align}
{^{(i)}}\!S_{s2}=\int\mathbb{S}_{\bf{k}_s}{^{(i)}}\!S_{tt}d^n\Omega,
\label{eq54.3}
\end{align}
\begin{align}
{^{(i)}}\!S_{s3}=\left(\frac{k_s^2}{nr}+\frac{(n-1)}{r}\right)\int^t{^{(i)}}\!S_{s1}dt-{^{(i)}}\!S_{s2}+\frac{n}{r}\left(\frac{r}{n}\int^t
{^{(i)}}\!S_{s1}dt\right)',
\label{eq54.4}
\end{align}
\begin{align}
{^{(i)}}\!S_{s4}=\frac{\int
\mathbb{S}^i_{\bf{k}_s}{^{(i)}}\!S_{ri}d^n\Omega}{k_s}+\left(\frac{{^{(i)}}\!S_{s0}}{k^2_s}\right)'-\left(\frac{{^{(i)}}\!S_{s0}}{rk^2_s}\right),
\label{eq54.5}
\end{align}
\begin{align}
{^{(i)}}\!S_{s5}=\int
\mathbb{S}_{\bf{k}_s}{^{(i)}}\!S_{rr}d^n\Omega+\frac{n}{rk^2_s}{^{(i)}}\!S_{s0}'-\frac{{^{(i)}}\!S_{s0}}{r^2},
\label{eq54.6}
\end{align}
\begin{align}
{^{(i)}}\!S_{s6}={^{(i)}}\!S_{s5}-\frac{n}{r}{^{(i)}}\!S_{s4},
\label{eq54.7}
\end{align}
\begin{align}
{^{(i)}}\!S_{s7}=\frac{{^{(i)}}\!S_{s3}}{2n}-\frac{(k_s^2-n)}{2n^2r}\int^t
{^{(i)}}\!S_{s1}dt+\frac{{^{(i)}}\!S_{s6}}{2n}-\frac{1}{2rk_s^2}\left[
r^2{^{(i)}}\!S_{s3}\right]',
\label{eq54.8}
\end{align}
Similar to the linear case, we need only five equations to get a master equation governing scalar perturbations at higher orders.
Firstly, we have equation (\ref{eq42}) which relates variables ${^{(i)}}\!F_{tt}$, ${^{(i)}}\!F_{rr}$ (through the trace ${^{(i)}}\!F_c^c$) and ${^{(i)}}\!F$:
\begin{align}
-k_s^2\left[2(n-2){^{(i)}}\!F+{^{(i)}}\!F^c_c\right]={^{(i)}}\!S_{s0}.
\label{eq55}
\end{align}
For $a=r,b=t$ in (\ref{eq40}) one gets
\begin{align}
\frac{n}{r}{^{(i)}}\!\dot{F}_{rr}+\frac{k_s^2}{r^2}{^{(i)}}\!F_{rt}-2n{^{(i)}}\!\dot{F}'-\frac{2n}{r}{^{(i)}}\!\dot{F}={^{(i)}}\!S_{s1}.
\label{eq56}
\end{align}
Now we write ${^{(i)}}\!F_{rt}$ in terms of the variable ${^{(i)}}\!\Psi_{s\textbf{k}_s}$ as
\begin{align}
{^{(i)}}\!F_{rt}=2r({^{(i)}}\!\dot{\Psi}_s+{^{(i)}}\!\dot{F}),
\label{eq58}
\end{align}
where ${^{(i)}}\!\Psi_s$ itself is defined in terms our master variable ${^{(i)}}\!\Phi_s$ as
\begin{align}
{^{(i)}}\!\Psi_s={^{(i)}}\!\Phi_s-{^{(i)}}\!S_{s8}.
\label{eq58.1}
\end{align}
The expression ${^{(i)}}\!S_{s8}$ is defined as
\begin{align}
{^{(i)}}\!S_{s8}=-\frac{1}{2}\left(\frac{k^2_s}{n}-1\right)r^{-(\frac{k_s}{\sqrt{n}}+n-1)}\int^r\left[r^{(\frac{2k_s}{\sqrt{n}}-1)}\int^r {r}'^{(-\frac{k_s}{\sqrt{n}}+n-2)}\:{^{(i)}}\!\mathcal{B} \;dr' \right]dr,
\label{eq58.2}
\end{align}
where
\begin{align}
{^{(i)}}\!\mathcal{B}(t,r)=\frac{n}{k^2_s-n}\left[r\left(\frac{r^2}{k_s^2}{^{(i)}}\!S_{s3}
\right)'  +\frac{r^2}{k^2_s}\left((n-1)-\frac{k_s^2}{n}  \right) {^{(i)}}\!S_{s3}\right]&+\frac{r}{n}\int^t{^{(i)}}\!S_{s1}dt\nonumber\\&+\frac{{^{(i)}}\!S_{s0}}{k_s^2}.
\label{eq58.3}
\end{align}
The above ansatz ensures that the boundary condition is devoid of  the products of the lower order metric perturbation contributed by the source terms. The details are given in appendix B.

We integrate (\ref{eq56}) with respect to $t$ and get an expression for
${^{(i)}}\!F_{rr}$ in terms of ${^{(i)}}\!F$ and ${^{(i)}}\!\Psi_s$ which is
\begin{align}
{^{(i)}}\!F_{rr}=2r{^{(i)}}\!F'+2{^{(i)}}\!F-\frac{2k_S^2}{n}{^{(i)}}\!F-\frac{2k^2_s}{n}{^{(i)}}\!\Psi_s+\frac{r}{n}\int^t
{^{(i)}}\!S_{s1}dt.
\label{eq59}
\end{align}
For $a=b=t$ in (\ref{eq40}) one gets
\begin{align}
-2n{^{(i)}}\!F''+\frac{n}{r}{^{(i)}}\!F_{rr}'+\Big(\frac{k_s^2}{r^2}&+\frac{n(n-1)}{r^2}\Big){^{(i)}}\!F_{rr}-\frac{2n(n+1)}{r}{^{(i)}}\!F'\nonumber\\&+\left(\frac{2k_s^2(n-1)}{r^2}-\frac{2n(n-1)}{r^2}\right){^{(i)}}\!F={^{(i)}}\!S_{s2}.
\label{eq60}
\end{align}
Substitution of (\ref{eq59}) in (\ref{eq60}) leads to an expression for ${^{(i)}}\!F$ in terms of ${^{(i)}}\!\Psi_s$ and its derivatives:
\begin{align}
{^{(i)}}\!F=-\frac{n}{k_s^2-n}\left[r{^{(i)}}\!\Psi_s'+\left(\frac{k_s^2}{n}+n-1\right){^{(i)}}\!\Psi_s-\frac{r^2}{2k_s^2}{^{(i)}}\!S_{s3}.
\right]
\label{eq61}
\end{align}
Consider the expansion of (\ref{eq41})  for $a=r$ in which we substitute
for ${^{(i)}}\!F_c^c$ from (\ref{eq55}). This gives:
\begin{align}
-\frac{(n-2)}{r}{^{(i)}}\!F_{rr}+{^{(i)}}\!\dot{F}_{rt}-{^{(i)}}\!F_{rr}'+2{^{(i)}}\!F'+\frac{2(n-2)}{r}{^{(i)}}\!F={^{(i)}}\!S_{s4}.
\label{eq63}
\end{align}
By making the substitution $a=b=r$ in (\ref{eq40}) and using (\ref{eq55}), one gets
\begin{align}
-2n{^{(i)}}\!\ddot{F}+\frac{2n}{r}{^{(i)}}\!F'-\frac{2k_s^2}{r^2}{^{(i)}}\!F&+\frac{2n(n-1)}{r^2}{^{(i)}}\!F-\frac{n}{r}{^{(i)}}\!F_{rr}'\nonumber\\&+\left(\frac{k_s^2}{r^2}-\frac{n(n-1)}{r^2}\right){^{(i)}}\!F_{rr}+\frac{2n}{r}{^{(i)}}\!\dot{F_{rt}}={^{(i)}}\!S_{s5}.
\label{eq66}
\end{align}
Now, we eliminate ${^{(i)}}\!F_{rr}'$ from (\ref{eq66}) by using (\ref{eq63}), to get
\begin{align}
-2n{^{(i)}}\!\ddot{F}+\frac{(k_s^2-n)}{r^2}{^{(i)}}\!F_{rr}-\frac{2(k_s^2-n)}{r^2}{^{(i)}}\!F+\frac{n}{r}{^{(i)}}\!\dot{F}_{rt}={^{(i)}}\!S_{s6}.
\label{eq68}
\end{align}
Substituting the expression for ${^{(i)}}\!F_{rt}$, ${^{(i)}}\!F_{rr}$ and ${^{(i)}}\!F$
from (\ref{eq58}), (\ref{eq59}) and (\ref{eq61}) in (\ref{eq68}) we obtain the following equation in terms of variable
${^{(i)}}\!\Phi_{s\textbf{k}_s}$
\begin{align}
{^{(i)}}\!\ddot{\Phi}_s+\hat{L}_s{^{(i)}}\!\Phi_s={^{(i)}}\!S_{s9},
\label{eq62}
\end{align}
where ${^{(i)}}\!S_{s9}$ is defined as
\begin{align}
{^{(i)}}\!S_{s9}={^{(i)}}\!S_{s7}+{^{(i)}}\!\ddot{S}_{s8}+\hat{L}_s{^{(i)}}\!S_{s8}.
\label{eq72}
\end{align}
We can now expand ${^{(i)}}\!\Phi_s$ in the basis of the eigenfunctions of the linear perturbation $e^{(s)}_{p,l_s}(r)$   as follows:
\begin{align}
{^{(i)}}\!\Phi_s=\sum_{p=0}^{\infty}{^{(i)}}\!c^{(s)}_{p,l_s}(t)e^{(s)}_{p,l_s}(r)
\label{eq73}
\end{align}
According to condition (\ref{eq74.11}), we require ${^{(i)}}\!F_{tt}$ to vanish at the boundary $r=\text{R}$, which implies ${^{(i)}}\!\Phi_{s\textbf{k}_s}$ should satisfy (see appendix B for further details)
\begin{align}
r^2{^{(i)}}\!\Phi_s''+(2n-1)r{^{(i)}}\!\Phi_s'+\left((n-1)^2-\frac{k_s^2}{n} \right){^{(i)}}\!\Phi_s=0\Big|_{r=\text{R}}
\label{eqe72.1}
\end{align}
The expansion (\ref{eq73}) ensures that this boundary condition is automatically satisfied --- this has been shown in appendix B.
One can  now use (\ref{add1}) to show that the ${^{(i)}}\!c_{q,l_s}^{(s)}(t)$
satisfy
\begin{align}
{^{(i)}}\!\ddot{c}_{q,l_s}^{(s)}+\omega^2_{q,l_s}{^{(i)}}\!c_{q,l_s}^{(s)}=<{^{(i)}}\!S_{s7},e_{q,l_s}^{(s)}>_s
\label{eq74}
\end{align}
where $<>_s$ is defined by  (\ref{add4}).

\section{Calculating the source terms}
The source terms ${^{(i)}}\!S_{\mu\nu}$ depend on ${^{(1)}}\!h_{\mu\nu}$, ${^{(2)}}\!h_{\mu\nu}$,...${^{(i-1)}}\!h_{\mu\nu}$. Hence, once we calculate ${^{(i)}}\!\Phi_{T\textbf{k}}$, ${^{(i)}}\!\Phi_{v\textbf{k}_v}$ and ${^{(i)}}\!\Phi_{s\textbf{k}_s}$, we need to use them to get back ${^{(i)}}\!h_{\mu\nu}$. Since we have chosen our gauge choice to be (\ref{eq24}), determining ${^{(i)}}\!f_{ab\textbf{k}_s}$, ${^{(i)}}\!f_{r\textbf{k}_s}^{(s)}$, ${^{(i)}}\!f_{a\textbf{k}_v}^{(v)}$ and ${^{(i)}}\!H_{T\textbf{k}}$, completely fixes the various components of ${^{(i)}}\!h_{\mu\nu}$.\\
\textbf{Tensor components}:\\
Since by definition, ${^{(i)}}\!H_{T\textbf{k}}={^{(i)}}\!\Phi_{\textbf{k}}$, determining $\Phi_{\textbf{k}}$ determines $H_{T\textbf{k}}$.\\
\textbf{Vector components}:\\
By definition, ${^{(i)}}\!Z_{r\textbf{k}_v}$ is related to ${^{(i)}}\!\Phi_{v\textbf{k}_v}$ through (\ref{eqe49.1l}) and (\ref{eq51}) for linear order and higher orders respectively.\\
${^{(i)}}\!Z_{t\textbf{k}_v}$ is related to ${^{(i)}}\!Z_{r\textbf{k}_v}$ through (\ref{eq47l}) and (\ref{eq48}) for linear and higher orders respectively.\\
Hence the vector components are given by ${^{(i)}}\!Z_{a}={^{(i)}}\!f_a^{(v)}$.\\
\textbf{Scalar components}:\\
Once the quantities ${^{(i)}}\!F_{\textbf{k}_s}$ and components of ${^{(i)}}\!F_{ab\textbf{k}_s}$ are determined in terms of ${^{(i)}}\!\Phi_{s\textbf{k}_s}$, the scalar components are given by:
\begin{align}
{^{(i)}}\!f_r=k_s{^{(i)}}\!F,
\end{align}
\begin{align}
{^{(i)}}\!f_{tt}={^{(i)}}\!F_{tt},
\end{align}
\begin{align}
{^{(i)}}\!f_{rr}={^{(i)}}\!F_{rr}-\frac{2}{k_s}(r{^{(i)}}\!f_r)'.
\end{align}
\begin{align}
{^{(i)}}\!f_{rt}={^{(i)}}\!F_{rt}-\frac{r}{k_s}{^{(i)}}\!\dot{f}_r
\end{align}

\section{Special modes}
\subsection{Scalar perturbations $l_s=0,1$ modes}
\subsubsection{$l_s=0$ mode}
In this case, $\mathbb{S}$ is constant and hence, $\mathbb{S}_i$ and
$\mathbb{S}_{ij}$ vanish. This means, only ${^{(i)}}\!f_{ab}$ and
${^{(i)}}\!H_L$ exist. We will use gauge freedom to put
\begin{align}
{^{(i)}}\!H_L={^{(i)}}\!f_{tt}=0
\label{eqss0}
\end{align}
Let ${^{(i)}}\!\tilde{S}_{0\:\mu\nu}=\int
\mathbb{S}_{l_s=0}{^{(i)}}\!S_{0\:\mu\nu}d^n\Omega$. We get the following
equations from  ${^{(i)}}\!G_{rt}=0$, ${^{(i)}}\!G_{tt}=0$ and
${^{(i)}}\!G_{rr}=0$ respectively.
\begin{align}
\frac{n}{r}{^{(i)}}\!\dot{f}_{rr}={^{(i)}}\!\tilde{S}_{0\:rt}
\label{eqss1}
\end{align}
\begin{align}
\frac{n}{r}{^{(i)}}\!{f}_{rr}'+\frac{n(n-1)}{r^2}{^{(i)}}\!f_{rr}={^{(i)}}\!\tilde{S}_{0\:tt}
\label{eqss2}
\end{align}
\begin{align}
\frac{2n}{r}{^{(i)}}\!\dot{f}_{rt}-\frac{n(n-1)}{r^2}{^{(i)}}\!f_{rr}={^{(i)}}\!\tilde{S}_{0\:rr}
\label{eqss3}
\end{align}
From (\ref{eqss1}), we can obtain ${^{(i)}}\!f_{rr}$ as:
\begin{align}
{^{(i)}}\!f_{rr}=\int_{t_1}^t\frac{r}{n}{^{(i)}}\!\tilde{S}_{0\:rt}dt+{^{(i)}}\!f_{rr}(t_1,r)
\label{eqss4}
\end{align}
 ${^{(i)}}\!f_{rr}(t_1,r)$ can be obtained from (\ref{eqss2}):
\begin{align}
{^{(i)}}\!f_{rr}(t_1,r)=\frac{1}{r^{n-1}}\int_0^r\frac{\bar{r}^n}{n}{^{(i)}}\!\tilde{S}_{0\:tt}(t_1,\bar{r})d\bar{r}
\label{eqss5}
\end{align}
Whereas, ${^{(i)}}\!f_{rt}$  is given by
\begin{align}
{^{(i)}}\!f_{rt}=\int^t\left[\frac{(n-1)}{2r}{^{(i)}}\!f_{rr}+\frac{r}{2n}{^{(i)}}\!\tilde{S}_{0\:rr}\right]dt
\label{eqss6}
\end{align}

\subsubsection{$l_s=1$ $(k_s^2=n)$ mode}
Let ${^{(i)}}\!\tilde{S}_{1\:\mu\nu}=\int \mathbb{S}_{l_s=1}{^{(i)}}\!S_{1\:\mu\nu}d^n\Omega$. Since $\mathbb{S}_{ij}$ vanishes for this mode, only ${^{(i)}}\!f_{ab}$, ${^{(i)}}\!f_a$ and ${^{(i)}}\!H_L$ exist. We will use gauge freedom to put ${^{(i)}}\!f_{tt}$, ${^{(i)}}\!f_t$ and ${^{(i)}}\!H_L$ to zero. Now we define the following quantities, composed solely of source terms.
\begin{align}
{^{(i)}}\!S_1=\frac{1}{r^n}\int_0^r\bar{r}^n\left[\frac{\bar{r}}{\sqrt{n}}{^{(i)}}\!\tilde{S}_{1\:tt}-\frac{1}{\sqrt{n}}\left(\bar{r}\int^t{^{(i)}}\!\tilde{S}_{1\:rt}dt\right)'-\sqrt{n}\left(\int^t{^{(i)}}\!\tilde{S}_{1\:rt}dt\right)\right]d\bar{r}
\label{eqs1}
\end{align}

\begin{align}
{^{(i)}}\!S_2=\left[\frac{1}{2\sqrt{n}}{^{(i)}}\!\tilde{S}_{1\:rr}+\frac{(n-1)}{2\sqrt{n}r}\int^t{^{(i)}}\!\tilde{S}_{1\:rt}dt+\frac{(n-1)}{2r^2}{^{(i)}}\!S_1
 \right]
\label{eqs2}
\end{align}

We will use the following four equations, namely ${^{(i)}}\!G_{rt}=0$, ${^{(i)}}\!G_{tt}=0$, ${^{(i)}}\!G_{rr}=0$ and ${^{(i)}}\!G_i^i=0$:
\begin{align}
\frac{n}{r}{^{(i)}}\!\dot{f}_{rr}+\frac{n}{r^2}{^{(i)}}\!f_{rt}+\frac{\sqrt{n}}{r}{^{(i)}}\!\dot{f}_r={^{(i)}}\!\tilde{S}_{1\:rt}
\label{eqs3}
\end{align}
\begin{align}
\frac{n}{r}{^{(i)}}\!f_{rr}'+\frac{n^2}{r^2}{^{(i)}}\!f_{rr}+\frac{2\sqrt{n}}{r}{^{(i)}}\!f_r'+\frac{n^{3/2}}{r^2}{^{(i)}}\!f_r={^{(i)}}\!\tilde{S}_{tt}
\label{eqs4}
\end{align}
\begin{align}
\frac{2n}{r}{^{(i)}}\!\dot{f}_{rt}-\frac{n(n-1)}{r^2}{^{(i)}}\!f_{rr}-\frac{2\sqrt{n}(n-1)}{r^2}{^{(i)}}\!f_r={^{(i)}}\!\tilde{S}_{rr}
\label{eqs5}
\end{align}
\begin{align}
{^{(i)}}\!\dot{f}_{rt}'+\frac{(n-1)}{r}{^{(i)}}\!\dot{f}_{rt}-\frac{1}{2}{^{(i)}}\!\ddot{f}_{rr}-\frac{(n-1)}{2r}{^{(i)}}\!f_{rr}'&-\frac{(n-1)}{\sqrt{n}r}{^{(i)}}\!f_r'-\frac{(n-1)^2}{2r^2}{^{(i)}}\!f_{rr}\nonumber\\&+\frac{(n-1)^2}{\sqrt{n}r^2}{^{(i)}}\!f_r=\frac{1}{n}{^{(i)}}\!\tilde{S}^i_{1\:i}
\label{eqs6}
\end{align}
We will redefine ${^{(i)}}\!f_{rt}$ as
\begin{align}
{^{(i)}}\!f_{rt}=\frac{r}{\sqrt{n}}{^{(i)}}\!\dot{\phi}_0
\label{eqs7}
\end{align}
Substituting this ansatz in (\ref{eqs3}) and then integrating w.r.t to $t$ gives,
\begin{align}
{^{(i)}}\!f_{rr}=-\frac{1}{\sqrt{n}}{^{(i)}}\!\phi_0-\frac{1}{\sqrt{n}}{^{(i)}}\!f_r+\int^t\frac{r}{n}
{^{(i)}}\!\tilde{S}_{1\:rt}dt
\label{eqs8}
\end{align}
The extra $r-$dependent integration function can be absorbed in the definition of ${^{(i)}}\!\phi_0$. Substituting the expression for ${^{(i)}}\!f_{rr}$ from (\ref{eqs8}) in (\ref{eqs4}) allows us to obtain ${^{(i)}}\!f_r$ in terms of ${^{(i)}}\!\phi_0$:
\begin{align}
{^{(i)}}\!f_r={^{(i)}}\!\phi_0+{^{(i)}}\!S_1
\label{eqs9}
\end{align}
Now, by substituting (\ref{eqs9}) in (\ref{eqs5}), one obtains:
\begin{align}
{^{(i)}}\!\ddot{\phi}_0={^{(i)}}\!S_2
\label{eqs10}
\end{align}
Hence from (\ref{eqs6}), we can obtain the following expression for ${^{(i)}}\!\phi_0$.

\begin{align}
{^{(i)}}\!\phi_0&=\frac{\sqrt{n}r^2}{2(n-1)^2}\Bigg[\frac{1}{n}{^{(i)}}\!\tilde{S}^i_{1\:i}+\frac{r}{2n}{^{(i)}}\!\dot{\tilde{S}}_{1\:rt}+\frac{(n-1)}{2r}\Big(\frac{r}{n}\int^t{^{(i)}}\!\tilde{S}_{1\:rt}dt\Big )'\nonumber\\& +\frac{(n-1)^2}{2nr}\int^t {^{(i)}}\!\tilde{S}_{1\:\!rt}dt+\frac{(n-1)}{2\sqrt{n}r}{^{(i)}}\!S_1'-\frac{3(n-1)^2}{2\sqrt{n}r^2}{^{(i)}}\!S_1\nonumber\\&-\frac{r}{\sqrt{n}}{^{(i)}}\!S_2'-\frac{(n+1)}{\sqrt{n}}{^{(i)}}\!S_2 -\frac{1}{2\sqrt{n}}\ddot{S}_1\Bigg]
\label{eqs11}
\end{align}

Once, ${^{(i)}}\!\phi_0$ is obtained, ${^{(i)}}\!f_{r}$, ${^{(i)}}\!f_{rr}$ and ${^{(i)}}\!f_{rt}$ can be determined using (\ref{eqs9}), (\ref{eqs8}) and (\ref{eqs7}) respectively.

\subsection{Vector perturbations $l_v=1$ $(k_v^2=n-1)$ mode}
Let ${^{(i)}}\!\tilde{S}^{(v)}_{1\:ia}=\int \mathbb{V}^i_{l_v=1}{^{(i)}}\!S_{1\:ia}d^n\Omega$ be the source associated with these modes. Since $\mathbb{V}_{ij}$ vanishes, only ${^{(i)}}\!f_a^{(v)}$ exist. Through a suitable gauge choice, one can put ${^{(i)}}\!f_t^{(v)}$ to zero. Thus, from ${^{(i)}}\!G_{ir}=0$ one can obtain ${^{(i)}}\!f_r^{(v)}$ as
\begin{align}
\partial_t {^{(i)}}\!f_r^{(v)}=\frac{1}{r}\int_{t_1}^t {^{(i)}}\!\tilde{S}_{1\:ir}^{(v)}dt'+{^{(i)}}\!\dot{f}_r^{(v)}(t_1,r)
\label{eqv1}
\end{align}
where ${^{(i)}}\!\dot{f}_r^{(v)}(t_1,r)$ is obtained from ${^{(i)}}\!G_{it}=0$ equation,
\begin{align}
{^{(i)}}\!\dot{f}_r^{(v)}(t_1,r)=\frac{1}{r^{n+1}}\int_0^r \bar{r}^n\:{^{(i)}}\!\tilde{S}^{(v)}_{1\:it}(t_1,\bar{r})d\bar{r}
\label{eqv2}
\end{align}

\section{Summary and Discussion}
In this article, we have analyzed perturbations of Minkowski spacetime with a spherical Dirichlet wall beyond linear order.
This is a model where it is possible to simplify the perturbation equations at arbitrary order, and the tools and techniques
we use can be generalized to study perturbations of $AdS$ spacetime. We work
in weakly nonlinear perturbation theory
and decompose the perturbations into scalar, vector and tensor spherical harmonics using the formalism of Ishibashi,
Kodama and Seto.
In contrast to $AdS$ where weakly nonlinear perturbation theory is expected to break down due to
irremovable secular terms,
we do not expect secular terms in this model. The system has already been shown to be stable at linear order \cite{marolf}.
Further, the spectrum for the linear tensor, scalar and vector
perturbations is asymptotically resonant as opposed to a resonant spectrum in the case of $AdS$ spacetime.
Even at linear order, the scalar sector of perturbations requires careful analysis and we
use techniques in \cite{soda} to analyze
the equations.
This is because the Dirichlet wall boundary conditions lead to a frequency-dependent boundary condition for the
scalar master variable (which depends on the scalar perturbations), a fact noted in \cite{marolf}.  Due to these boundary conditions, the scalar eigenfunctions are not orthogonal
with respect to the usual inner product. We define a modified orthogonality relation which the eigenfunctions satisfy.  Going beyond linear order, by fixing gauge appropriately, we present the (nonhomogeneous) perturbation
equations at arbitrary order in a simplified form. The source terms are made of lower order perturbations. At any order,
the perturbation consists of scalar, vector and tensor-type parts. The equation for each of these is derived by projecting onto the
space of perturbations of each type. Once these equations are obtained, we analyze each type separately at arbitrary order. The
tensor perturbations are straightforward to analyze. The perturbation at arbitrary order can be written in terms of the eigenbasis of
linear tensor perturbations with time dependent coefficients. These time dependent coefficients satisfy a simple
forced harmonic oscillator equation. In the case of vector and scalar perturbations, we define new shifted master variables
(shifted by source terms) such that these new variables obey the same boundary conditions as the linear perturbations. They are thus
expanded in an eigenbasis of linear perturbations with time dependent coefficients satisfying a forced harmonic oscillator equation.
This (forced harmonic oscillator) structure of the equations implies that we can apply the nonlinear dynamics analysis in \cite{ads}
that uses Hamiltonian perturbation theory. The system is described by a Hamiltonian that is a perturbation of the integrable
Hamiltonian of
linear
harmonic oscillators. This analysis explained the result of previous numerical studies of the Einstein-scalar field system where
an instability was seen in a cavity with both Dirichlet and Neumann boundary conditions. Neumann boundary conditions resulted in
an asymptotically resonant spectrum, yet an instability was seen \cite{pani}. However, a certain minimum amplitude of the scalar field was required to trigger instability in the Neumann case.
In particular, for our system of gravitational perturbations with a spherical Dirichlet wall,
the asymptotically resonant spectrum for linear order perturbations
implies that the system is stable under arbitrarily small perturbations but that instability could set in for
perturbations whose magnitude depends on a number-theoretic measure of the deviation of the spectrum from a resonant one. This may also be possible
to see numerically. Finally, we analyze certain special modes separately. These are the scalar modes with $l_s = 0$, $l_s = 1$ and the vector
mode with $l_{v} = 1$ for which the equations become gauge dependent. By a choice of gauge fixing, we analyze these perturbations
at arbitrary order. It is possible to integrate the equations and write the form of the
perturbations. As expected, at linear order, these are pure gauge.

One of the interesting questions we have not addressed and indeed, can be answered only numerically is the fate of the system for gravitational perturbations of appropriate magnitude that may trigger instability --- whether a rotating black hole is the result.

\section{Appendix A}
We use (\ref{eq4}) and (\ref{eq5}) to get the expansions of metric perturbation $\delta g _{\mu\nu}$ and Christoffel's symbol to second order. In the following the superscript on the left hand side of a quantity denotes the order of $\epsilon$.
\begin{align}
\delta\Gamma_{\mu\nu}^{\alpha}={^{(1)}}\Gamma_{\mu\nu}^{\alpha}+{^{(2)}}\Gamma_{\mu\nu}^{\alpha}...
\label{eqa1}
\end{align}
where
\begin{align}
{^{(1)}}\Gamma_{\mu\nu}^{\alpha}=\frac{1}{2}\left(  \bar{\nabla}_{\mu}h^{\alpha}_{\nu}+\bar{\nabla}_{\nu}h^{\alpha}_{\mu}-\bar{\nabla}^{\alpha}h_{\mu\nu}\right)
\label{eqa2}
\end{align}
\begin{align}
{^{(2)}}\Gamma^{\alpha}_{\mu\nu}=\frac{1}{2}\left( \bar{\nabla}_{\mu}{^{(2)}}h^{\alpha}_{\nu}+\bar{\nabla}_{\nu}{^{(2)}}h^{\alpha}_{\nu}-\bar{\nabla}^{\alpha}{^{(2)}}h_{\mu\nu}\right)-\frac{1}{2}h^{\alpha\lambda}\left( \bar{\nabla}_{\mu}h_{\lambda\nu}+\bar{\nabla}_{\nu}h_{\lambda\mu} -\bar{\nabla}_{\lambda}h_{\mu\nu}\right)
\label{eqa3}
\end{align}
\begin{align}
\delta R_{\mu\nu}={^{(1)}}R_{\mu\nu}+{^{(2)}}R_{\mu\nu}...
\label{eqa4}
\end{align}
where
\begin{align}
{^{(1)}}R_{\mu\nu}=\triangle_L h_{\mu\nu}=0
\label{eqa5}
\end{align}
\begin{align}
2{^{(2)}}R_{\mu\nu}=&2\triangle_L {^{(2)}}h_{\mu\nu}+\frac{1}{2}\bar{\nabla}_{\alpha}h\left(   -\bar{\nabla}^{\alpha}h_{\mu\nu}+\bar{\nabla}_{\nu}h^{\alpha}_{\mu}+\bar{\nabla}_{\mu}h^{\alpha}_{\nu} \right)\nonumber\\&-h^{\lambda\alpha}\bigg(  -\bar{\nabla}_{\lambda}\bar{\nabla}_{\alpha}h_{\mu\nu}-\bar{\nabla}_{\mu}\bar{\nabla}_{\nu}h_{\lambda\alpha} +\bar{\nabla}_{\lambda}\bar{\nabla}_{\mu}h_{\alpha\nu}+\bar{\nabla}_{\alpha}\bar{\nabla}_{\nu}h_{\mu\lambda} \bigg)\nonumber\\&+\frac{ \bar{\nabla}_{\nu}h^{\lambda\alpha}\bar{\nabla}_{\mu}h_{\lambda\alpha}}{2}-\bar{\nabla}_{\lambda}h^{\alpha}_{\mu}\bar{\nabla}_{\alpha}h^{\lambda}_{\nu}+\bar{\nabla}^{\alpha}h_{\mu\lambda}\bar{\nabla}_{\alpha}h^{\lambda}_{\nu}\nonumber\\&-\bar{\nabla}_{\lambda}h^{\lambda\alpha}\bigg(-\bar{\nabla}_{\alpha}h_{\mu\nu}+\bar{\nabla}_{\nu}h_{\mu\alpha}+\bar{\nabla}_{\mu}h_{\alpha\nu}\bigg)
\label{eqa6}
\end{align}
\begin{align}
{^{(2)}}\!R=&\bar{g}^{\mu\nu}\:{^{(2)}}\!R_{\mu\nu}-h^{\mu\nu}\:{^{(1)}}\!R_{\mu\nu}\nonumber\\=&\bar{g}^{\mu\nu}\:{^{(2)}}\!R_{\mu\nu}
\label{eqa7}
\end{align}
The second order in $\epsilon$ Einstein's equation is
\begin{align}
2{^{(2)}}\!R_{\mu\nu}-{^{(2)}}\!h_{\mu\nu}\bar{R}-\bar{g}_{\mu\nu}{^{(2)}}\!R-h_{\mu\nu}{^{(1)}}\!R=0
\label{eqa8.1}
\end{align}
where the last term vanishes because of condition (\ref{eqa5}).
Substituting the appropriate expressions in (\ref{eqa8.1}) one finds the second order equation to be
\begin{align}
\tilde{\triangle}_L{^{(2)}}h_{\mu\nu}={^{(2)}}S_{\mu\nu}
\label{eqa8}
\end{align}
\section{Appendix B}
From using (\ref{eq59}) and (\ref{eq55}) we obtain the following expression for ${^{(i)}}\!F_{tt}$:
\begin{align}
{^{(i)}}\!F_{tt}=2r{^{(i)}}\!F'+2(n-1){^{(i)}}\!F-\frac{2k_s^2}{n}{^{(i)}}\!F-\frac{2k_s^2}{n}{^{(i)}}\!\Psi+\frac{r}{n}\int^t
{^{(i)}}\!S_{s1}dt+\frac{{^{(i)}}\!S_{s0}}{k_s^2}
\label{eqb1}
\end{align}
Now we substitute for ${^{(i)}}\!F$ and ${^{(i)}}\!\Psi_s$ from (\ref{eq61}) and (\ref{eq58.1}) in the above equation
\begin{align}
{^{(i)}}\!F_{tt}=&-\frac{n}{k_s^2-n}\left[2r^2{^{(i)}}\!\Phi_s''+2r(2n-1){^{(i)}}\!\Phi_s'+2\left(
(n-1)^2-\frac{k^2_s}{n}\right){^{(i)}}\!\Phi_s\right]\nonumber\\&+\frac{n}{k_s^2-n}\left[2r^2{^{(i)}}\!S_{s8}''+2r(2n-1
){^{(i)}}\!S_{s8}'+2\left( (n-1)^2-\frac{k_s^2}{n} \right){^{(i)}}\!S_{s8}
\right]+{^{(i)}}\!\mathcal{B}
\label{eqb2}
\end{align} where ${^{(i)}}\!\mathcal{B}$ is given by (\ref{eq58.3}).
We wish to choose a form for ${^{(i)}}\!S_{s8}$ which will ensure that the
terms in second line of (\ref{eqb2}) vanish.
Define
 \begin{align}
 {^{(i)}}\!S_{s8}={^{(i)}}\!\chi f
 \label{eqb2.1}
 \end{align} where
\begin{align}
f=r^{-(\frac{k_s}{\sqrt{n}}+n-1)};~~{^{(i)}}\!\chi=-\frac{k_s^2-n}{2n}\int^r\left[r^{(\frac{2k_s}{\sqrt{n}}-1)}\int^r {r'}^{(-\frac{k_s}{\sqrt{n}}+n-2)}\:{^{(i)}}\!\mathcal{B} \;dr' \right]dr
\label{eqb3}
\end{align}
Substitute (\ref{eqb2.1}) in (\ref{eqb2}) one obtains
\begin{align}
{^{(i)}}\!F_{tt}=&-\frac{n}{k_s^2-n}\left[2r^2{^{(i)}}\!\Phi_s''+2r(2n-1){^{(i)}}\!\Phi_s'+2\left(
(n-1)^2-\frac{k^2_s}{n}\right){^{(i)}}\!\Phi_s\right]\nonumber\\&
+\frac{2n}{k_s^2-n}\Bigg\{{^{(i)}}\!\chi\left[r^2f''+r(2n-1)f'+\left((n-1)^2-\frac{k_s^2}{n}\right)f\right]\nonumber\\&+r^2f\left[
{^{(i)}}\!\chi''+{^{(i)}}\!\chi'\left( \frac{(2n-1)}{r}+2\frac{f'}{f}
\right)\right]\Bigg\}+{^{(i)}}\!\mathcal{B}
\label{eqb4}
\end{align}
One can easily see that for the  choice of $f$ and ${^{(i)}}\!\chi$ given by (\ref{eqb3}), the last two lines in (\ref{eqb4}) vanish.
Hence the expression for ${^{(i)}}\!F_{tt}$ is
\begin{align}
{^{(i)}}\!F_{tt}=&-\frac{n}{k_s^2-n}\left[2r^2{^{(i)}}\!\Phi_s''+2r(2n-1){^{(i)}}\!\Phi_s'+2\left((n-1)^2-\frac{k^2_s}{n}\right){^{(i)}}\!\Phi_s\right]
\label{eqb5}
\end{align}
Applying  Dirichlet condition (\ref{eq74.11}), then gives

\begin{align}
r^2{^{(i)}}\!\Phi_s''+r(2n-1){^{(i)}}\!\Phi_s'+\left((n-1)^2-\frac{k^2_s}{n}\right){^{(i)}}\!\Phi_s=0\Bigg|_{r=\text{R}}
\label{eqb6}
\end{align}
The expansion (\ref{eq73}) ensures that condition (\ref{eqb6}) is automatically satisfied. This can be seen as follows:\\ In terms of expansion (\ref{eq73}), ${^{(i)}}\!F_{tt}$ is
\begin{align}
{^{(i)}}\!F_{tt}=-\frac{2n}{k_s^2-n}\sum_{p=0}^{\infty}{^{(i)}}\!c^{(s)}_{p,l_s}\left[r^2e_{p,l_s}^{(s)}\,\!''+(2n-1)re_{p,l_s}^{(s)}\,\!' +\left((n-1)^2-\frac{k_s^2}{n}\right)e_{p,l_s}^{(s)} \right].
\label{eqb8}
\end{align}
Since $e^{(s)}_{p,l_s}$ satisfy (\ref{eq74l}), $r^2e_{p,l_s}^{(s)}\,\!''=(-r^2\omega^2+k_s^2)e_{p,l_s}^{(s)}-nre_{p,l_s}^{(s)}\,\!'$.  Hence by the use of this expression in (\ref{eqb8}) one obtains:
\begin{align}
{^{(i)}}\!F_{tt}=\sum_{p=0}^{\infty}\frac{2n{^{(i)}}\!c^{(s)}_{p,l_s}}{n-k_s^2}\left[ (n-1)re_{p,l_s}^{(s)}\,\!'+\left(-\omega^2r^2+\frac{(n-1)}{n}(k_s^2+n(n-1)) \right)e^{(s)}_{p,l_s} \right],
\label{eqb9}
\end{align}
which vanishes at $r=\text{R}$ because of (\ref{eq73.2l}).
\section{Appendix C}
The expansion of source terms in general is tedious. Nevertheless, here we give an example by considering a simple case. Suppose we start out with only tensor-type perturbations at the linear level. Then ${^{(2)}}\!A_{ij}$ is given by:
\begin{align}
{^{(2)}}\!A_{ij}=&\sum_{\bf{k_1}}  \sum_{\bf{k_2}}H_{T\bf{k_1}}H_{T\bf{k_2}}\Big(T^{kl}_{\bf{k_1}}(-D_iD_jT_{kl_{\bf{k_2}}}+D_kD_iT_{jl_{\bf{k_2}}}+D_kD_jT_{li_{\bf{k_2}}}
\nonumber\\&-D_kD_lT_{ij_{\bf{k_2}}})-\frac{D_iT^{kl}_{\bf{k_1}}D_jT_{kl_{\bf{k_2}}}}{2}+D_kT^l_{i_{\bf{k_1}}}D_lT^k_{j_{\bf{k_2}}}-D^kT_{il_{\bf{k_1}}}D_kT^l_{j_{\bf{k_2}}}\Big)\nonumber\\&
-rD^arD_aH_{T\bf{k_1}}H_{T\bf{k_2}}\gamma_{ij} T^{kl}_{\bf{k_1}}T_{kl_{\bf{k_2}}}-r^2D^aH_{T\bf{k_1}}D_aH_{T\bf{k_2}}T_{ik_{\bf{k_1}}}T^k_{j_{\bf{k_2}}}
\end{align}
Similarly, ${^{(2)}}\!A_{ai}$ and ${^{(2)}}\!A_{ab}$ are,
\begin{align}
{^{(2)}}\!A_{ai}=&\sum_{\textbf{k}_1,\textbf{k}_2}\Bigg\{ H_{T\textbf{k}_1}\bar{D}_aH_{T\textbf{k}_2}\mathbb{T}^{kl}_{\textbf{k}_1}\bigg(-\bar{D}_i\mathbb{T}_{\textbf{k}_2kl}+\bar{D}_k\mathbb{T}_{\textbf{k}_2il}\bigg)\nonumber\\&-\frac{1}{2}\bar{D}_a H_{T\textbf{k}_1}H_{T\textbf{k}_2}\mathbb{T}^{kl}_{\textbf{k}_1}\bar{D}_i\mathbb{T}_{\textbf{k}_2kl}\Bigg\}
\end{align}
\begin{align}
{^{(2)}}\!A_{ab}=&\sum_{\textbf{k}_1,\textbf{k}_2}\Bigg\{ \bigg(-H_{T\textbf{k}_1}\bar{D}_a\bar{D}_bH_{T\textbf{k}_2}-\bar{D}_aH_{T\textbf{k}_1}\bar{D}_bH_{T\textbf{k}_2}\nonumber\\&-\frac{1}{r}\bar{D}_arH_{T\textbf{k}_1}\bar{D}_bH_{T\textbf{k}_2}-\frac{1}{r}\bar{D}_br\bar{D}_aH_{T\textbf{k}_1}H_{T\textbf{k}_2}\bigg)\mathbb{T}^{ij}_{\textbf{k}_1}\mathbb{T}_{\textbf{k}_2ij}  \Bigg\}
\end{align}

\end{document}